\documentclass[a4paper, 11pt]{article}
\pdfoutput=1
\usepackage{jcappub} 
\usepackage{lineno}
\usepackage{graphicx}
\usepackage{amsmath}
\usepackage{mathtools}
\usepackage{microtype}
\usepackage{cancel}
\usepackage{bigints}
\usepackage{bm}
\usepackage[usenames, dvipsnames]{color}
\usepackage{soul}
\usepackage{csquotes}
\usepackage{subcaption}
\usepackage{enumitem}
\usepackage{xspace}
\usepackage{booktabs}
\usepackage{siunitx}
\usepackage{ragged2e}
\usepackage{tensor}
\usepackage[normalem]{ulem}
\usepackage{cleveref}
\usepackage{hyperref}
\usepackage{feynmp-auto}
\usepackage{mleftright}
\makeatletter
\input{aas_macros.sty}
\let\jnl@style=\relax
\makeatother

\DeclareSIUnit{\year}{yr}
\DeclareSIUnit{\pc}{pc}
\DeclareSIUnit{\kpc}{\kilo\pc}
\DeclareSIUnit{\Mpc}{\mega\pc}
\DeclareSIUnit{\cLight}{\text{\ensuremath{c}}}
\DeclareSIUnit{\hHubble}{\text{\ensuremath{h}}}
\DeclareSIUnit{\Msun}{\text{\ensuremath{M_\odot}}}

\newcommand*{\ddelta}{{\mathchar '26\mkern -10mu\delta_{\mathrm{D}}}}
\newcommand{\bx}{{\bm x}}

\newcommand{\bk}{{\bm k}}
\newcommand{\bp}{{\bm p}}

\newcommand{\bv}{{\bm v}}
\newcommand{\ba}{{\bm a}}

\newcommand{\beq}{\begin{equation}\begin{aligned}\relax}
\newcommand{\eeq}{\end{aligned}\end{equation}}

\newcommand*{\ml}{\mleft}
\newcommand*{\mr}{\mright}
\newcommand{\mT}[2]{\mathcal{T}^{(\mathrm{#1})}_{#2}}
\newcommand{\mTi}{\mathcal{T}^{(i)}_k}
\newcommand{\mTfi}{\mathcal{T}^{{\rm fs}\,(i)}_k}
\newcommand{\mTf}[2]{\mathcal{T}^{{\rm fs}\,(\mathrm{#1})}_{#2}}

\newcommand{\Mpc}{{\rm Mpc}}
\newcommand{\eV}{{\rm eV}}

\newcommand{\di}{\mathrm{d}}
\newcommand{\dl}{\mathrm{d}}
\usepackage[most]{tcolorbox}

\newtcolorbox{eqnbox}{
  colback=gray!10,    
  colframe=gray!80,   
  arc=4mm,            
  boxrule=0.5mm,      
  left=2mm, right=2mm, top=1mm, bottom=1mm  
}

\newcommand{\ma}[1]{{\textcolor{Black}{{#1}}}}

\newcommand{\sd}[1]{{\textcolor{Black}{{#1}}}}

\definecolor{green2}{cmyk}{0.27, 0, 1, 0.52}
\hypersetup{
    colorlinks,            
    linkcolor=green2,      
    citecolor=green2,      
    filecolor=magenta,     
    urlcolor=green2        
}

\title{Multi-species Dark Matter with Warmth and Randomness}
\author[a]{Mustafa A. Amin}
\emailAdd{mustafa.a.amin@rice.edu}
\author[b]{, M. Sten Delos}
\emailAdd{mdelos@carnegiescience.edu}
\author[a]{, Kaixin Yang}
\emailAdd{ky35@rice.edu}

\affiliation[a]{Department of Physics and Astronomy, Rice University,
Houston, TX 77005, USA}
\affiliation[b]{Carnegie Observatories, 813 Santa Barbara Street, Pasadena, CA 91101, USA}

\abstract{We present a general analytic framework for the evolution of cosmic structure in multi-species dark matter models that simultaneously incorporates finite velocity dispersion and Poisson fluctuations.  Our approach accommodates arbitrary numbers of dark matter components with distinct mass fractions, velocity distributions, and number densities -- ranging from cold particles to warm species and sparse populations such as primordial black holes or solitons.  The framework is based on solving a truncated BBGKY hierarchy, whose solution is obtained by solving Volterra integral equations. We provide an efficient algorithm to solve for the total, as well as inter- and intra-species power spectra. Worked examples with two-component mixtures illustrate how isocurvature (initially Poisson) and adiabatic spectra evolve differently depending on the properties of the warm or sparse fraction. This evolution is controlled by the free-streaming and Jeans scales, and the results match analytic estimates and $N$-body simulations.}

\begin{document}
\maketitle
\flushbottom
\section{Introduction}
\label{sec:intro}
Evidence for dark matter (DM), a non-relativistic, gravitationally interacting component that dominates the mass density of our universe, has grown steadily over the past nine decades \cite{Cirelli:2024ssz,Marsh:2024ury}. Yet, its production mechanism, particle mass, spin, 
and interaction properties
remain unknown. We also do not know how complex the dark sector is. The dark sector could consist of a single species, or be a multi-species system. Each species may have distinct initial conditions and microscopic properties, while coupling to the others at least gravitationally. In this paper we develop a framework for evolving the density perturbations in such a multi-species system.

On cosmological and astrophysical scales, visible matter observations require DM to cluster under its own gravity. This clustering is quantified by the matter density contrast power spectrum (PS), whose shape and evolution encode information about the dark sector. Observations tightly constrain the PS on comoving scales $\gtrsim\rm Mpc$, showing behavior consistent with nearly scale-invariant adiabatic initial conditions and scale-independent growth of perturbations on subhorizon scales. Such results are compatible with many DM models. However, deviations from this simple behavior -- such as scale-dependent suppression or enhancement -- could reveal the microscopic properties and production history of DM \cite{Drlica-Wagner:2022lbd}. Significant observational effort is being dedicated to look for such departures \cite{Mondino:2020rkn,Sabti:2021unj,Gilman:2021gkj,Delos:2021ouc,Drlica-Wagner:2022lbd,Boylan-Kolchin:2022kae,Chung:2023syw,Irsic:2023equ,Delos:2023dwq,Esteban:2023xpk,Nadler:2024ims,An:2025gju,Xiao:2024qay,Ji:2024ott,deKruijf:2024voc,Buckley:2025zgh,Boddy:2025oxn,Trost:2024ciu,Fernandez:2023ddy}.

A general deviation from scale invariance is a Poisson, or white-noise, component on small scales, which arises naturally when the DM constituents have a sufficiently low number density. The non-observation of this feature has constrained the number density of the dominant dark matter component to $\bar{n} \gtrsim 10^8\,{\rm Mpc}^{-3}$.
These constituents are not elementary particles; they could be composite objects such as primordial black holes (PBHs) \cite{Zeldovich:1967pbh,Afshordi:2003zb,Inman:2019wvr,Carr:2023tpt,Delos:2024poq,Ivanov:2025pbu,Gerlach:2025vco}, interference granules in wave dark matter \cite{Schive:2014dra,Hui:2016ltb,Eberhardt:2025caq}, solitons \cite{Kaup:1968boson,Khlopov:1985jw,Chavanis:2011zm,Schive:2014dra,Zhang:2024bjo,Zhou:2024mea}, miniclusters and minihalos \cite{Kolb:1993zz,Hogan:1988mp,Delos:2017thv,Delos:2018ueo,Eggemeier:2019khm,StenDelos:2022jld,Delos:2023fpm,Gorghetto:2024vnp}, etc. 

Another generic deviation arises from warmth of the dark matter species. For thermally produced DM, if the particle is sufficiently light, free streaming during radiation domination suppresses the otherwise nearly scale-invariant spectrum on small scales. The absence of this suppression has placed an upper bound on the particle mass of order a few keV (e.g.~\cite{Bode:2000gq,Viel:2005qj,Viel:2013apy,Irsic:2023equ}). Without assuming thermal production, the bound relaxes to $m\gtrsim 10^{-19}\,\eV$ \cite{Amin:2022nlh}. In the thermal and non-thermal cases, the bound corresponds to an effective velocity dispersion $\lesssim 10\,{\rm km}\,{\rm s}^{-1}$ at matter–radiation equality. 

More generally, significant velocity dispersion (``warmth'') along with adiabatic and Poisson fluctuations can coexist. For example, ultralight bosonic DM produced after inflation (e.g. \cite{Petrossian-Byrne:2025mto}) is expected to exhibit significant velocity dispersion and Poisson fluctuations (in addition to usual adiabatic ones), due to causality considerations \cite{Amin:2022nlh}. \sd{Meanwhile, primordial black holes as dark matter naturally have significant Poisson fluctuations in their number density, and they can also be produced with significant velocities (e.g.~ \cite{Jenkins:2020ctp,Yuwen:2024gcf,Wang:2025mea})}. In recent work, some of us developed an analytic framework (validated by numerical simulations) describing how such adiabatic and Poisson fluctuations in single-species particle and wave DM evolve in the presence of velocity dispersion across radiation- and matter-dominated eras \cite{Amin:2022nlh,Ling:2024qfv,Amin:2025dtd,Amin:2025sla}. For related recent work on wave DM, also see \cite{Liu:2024pjg, Long:2024imw,  Liu:2025lts,Harigaya:2025pox,Ling:2025ehm}. 

Here, we extend our framework to multi-species dark matter with general initial phase-space distributions. This approach encompasses a broad range of scenarios, from mixtures of cold or warm components with adiabatic initial conditions to populations of PBHs, solitons, or wave-interference structures that generate Poisson fluctuations with or without velocity dispersion. Multi-species dark sectors are common in many high energy physics models \cite{Zurek:2008qg,Arvanitaki:2009fg,Cicoli:2023opf,Alexander:2024nvi}. Moreover, even if the fundamental particles are a single species, they might exist in different phases, such as a subdominant fraction in solitons, miniclusters, interference granules, or locked into PBHs. 

Even a subdominant component $S$ with mass fraction $\mathfrak{f}_S\ll 1$ can imprint a measurable white-noise floor in the power spectrum of order $\mathfrak{f}_S^2/\bar{n}_S$ if it has a sufficiently low number density and, via gravitational coupling, seed perturbations in the dominant component. If such a species is warm, the shape of the Jeans suppression in its own spectrum can be different from that in the dominant species. Moreover, the lack of clustering due to warmth (even in absence of white noise), can lead to well known shallow suppressions of order $\mathfrak{f}_S$ in the adiabatic power spectrum. 

Following our earlier work \cite{Amin:2025dtd}, the calculation is built on a truncated BBGKY (Bogoliubov–Born–Green–Kirkwood–Yvon) hierarchy, now generalized to multiple species. The resulting solution expresses the power spectrum in terms of three families of transfer functions governed by Volterra integral equations. To improve readability of our manuscript, we present the detailed derivation in Appendix \ref{App:Derivation} and only provide the key results for the evolving power spectrum in the main text (Sec.~\ref{sec:Main}). \ma{The main results are followed by representative multi-component examples (Sec.~\ref{sec:Examples}), including validation with $N$-body simulations. A numerical algorithm and publicly available code for evaluating the power spectrum evolution is provided in Appendix \ref{App:Num-PS}.}

We note that this paper focuses on multi-species dark matter treated as classical point particles. A companion paper \cite{Amin:2025nxm} addresses multi-species wave dark matter, where additional de Broglie-scale wave effects are included.

Understanding growth of structure in mixed and multi-species dark matter has a long history (e.g.~\cite{Primack1997,Schwabe:2020eac,Cyncynates:2021xzw,Vanzan:2023gui,Lague:2023wes,Tan:2024cek,Garcia-Gallego:2025kiw}). In particular, including warmth in a fraction of the dark matter has been explored extensively, including in the context of neutrinos (e.g~\cite{Lewis:2002nc,Lesgourgues:2012uu}). Recently, the EFT of large scale structure formalism has been applied to such scenarios \cite{Celik:2025wkt,Verdiani:2025jcf}. Poisson noise due to primordial black holes or miniclusters has also been considered in the past (eg.~\cite{Afshordi:2003zb,Zurek:2006sy,Carr:2018rid,McQuinn}). What is new in this work is the ability to deal with discreteness effects (Poisson fluctuations) and warmth together in some fraction of dark matter, and its impact on the rest of the dark matter species. The standard non-Poisson adiabatic perturbation evolution is also naturally included. The framework we develop is general enough to include an arbitrary number of species with varying fractions, warmth, and shot-noise contributions. \ma{As with standard perturbation theory, our framework is restricted to the linear evolution of the density perturbations.}

\section{Model \& Main Results}
\label{sec:Main}
We suppose that dark matter is composed of $\mathcal{N}$ species of particles. Each species, labeled by $``S"$, is made up of particles of mass $m_S$, with mass and number density $n_S(\bx)$ and $\rho_S(\bx)=m_Sn_S(\bx)$ respectively. Their spatially averaged counterparts are denoted by $\bar{n}_S$ and $\bar{\rho}_S$. The total dark matter density $\rho=\sum_{S=1}^\mathcal{N}{\rho}_S$. Each species has its own velocity distribution $f^{S}(\bv)$, and a velocity dispersion $\sqrt{\langle \bv^2\rangle} \ll c$. The ``particles" can be fundamental or composite, they can be solitons, PBHs and even quasi-particles associated with wave dark matter. We treat them all as classical point particles.

We will assume statistical homogeneity and isotropy. The background expansion of the universe is determined by a radiation and dark matter energy density, with the Hubble parameter $H(y)=({k_{\mathrm{eq}}}/{\sqrt{2}a_{\mathrm{eq}}})y^{-2}\sqrt{1+y}$, where $y=a/a_{\rm eq}$.
The scale factor $a$ at matter–radiation equality is $a_{\mathrm{eq}}\approx 1/3388$, and the comoving wavenumber associated with the horizon size at that time is $k_{\mathrm{eq}} = a_{\mathrm{eq}}H(a_{\mathrm{eq}})\approx \SI{0.01}{\per\Mpc}$ \cite{Planck:2018vyg}. We restrict our attention to subhorizon scales.

Our goal is to understand the growth of density perturbations in dark matter during radiation and matter domination. To this end, we define the density contrast in each species $\delta_S(\bx)=[\rho_S(\bx)-\bar{\rho}_S]/\bar{\rho}_S$, with the total density contrast $\delta(\bx)= [\rho(\bx)-\bar{\rho}]/\bar{\rho}=\sum_S \mathfrak{f}_S \delta_S(\bx)$ where $\mathfrak{f}_S\equiv \bar{\rho}_S/\bar{\rho}$. Their evolution is determined once we specify the initial velocity distributions $f^S(\bv)$ for each species at some early time in the radiation era, once all the species are non-relativistic.

In Appendix \ref{App:Derivation}, starting with the Liouville equation for this gravitationally clustering multispecies system, we arrive at expressions for the time evolution of the power spectrum of this density contrast. This derivation is similar to the one presented in Ref.~\cite{Amin:2025dtd} which assumed a single species. 
\sd{It neglects 3-particle and higher-order correlations, so the results are valid at linear order in perturbations, as discussed in Ref.~\cite{Amin:2025dtd}.}
In this work, the derivation is generalized to multiple species. The main results are as follows.

\subsection{Total Power Spectrum}
The time evolution of the power spectrum of the total density contrast is:
\beq
    \label{eq:MainResultPS}
    P_{\delta}(y,k) &= \underbrace{{P}^{(\mathrm{ad})}_{\delta}(y_0,k) \ml[\mathcal{T}_k^{(\mathrm{ad})}(y,y_0)\mr]^2 \vphantom{\int_{y_0}^y}}_{\text{adiabatic IC + evolution}} \!\! +
    \underbrace{P^{(\mathrm{iso})}_\delta(y_0,k) \ml[\mathcal{T}_k^{(\rm iso)}(y,y_0)\mr]^2}_{\text{isocurvature IC + evolution}}
\eeq
where the adiabatic and isocurvature transfer functions\footnote{In our context, the Poisson contribution is generated post inflation and is isocurvature in nature. It is uncorrelated with the adiabatic initial conditions from inflation.} are given by
\beq \label{eq:MainResultT}
\mathcal{T}_k^{({\rm iso})}(y,y_0) &=
 \ml[1 + 3 \int_{y_0}^y \! \frac{\dl y'}{\sqrt{1+y'}} \mathcal{T}^{(\mathrm{b})}_k(y, y')\mathcal{T}^{(\mathrm{c})}_k(y,y')\mr]^{1/2}\,,\\
\mathcal{T}^{(\mathrm{ad})}_k(y,y_0) &= \mathcal{T}^{(\mathrm{a})}_k(y,y_0) + \frac{1}{2} \frac{\dl\ln(P^{({\mathrm{ad}})}_{\delta}(y_0,k))}{\dl\ln(y_0)} \sqrt{1+y_0} \, \mathcal{T}^{(\mathrm{b})}_k(y,y_0)\,.
\eeq
Here, $y_0\ll 1$ is at an initial ``time" when all wavenumber-$k$ modes of interest are sub-horizon, and the field modes of interest are non-relativistic; the initial conditions (IC) are specified at that time. Note that isocurvature ``initial condition", $P_\delta^{(\rm iso)}(y_0,k)=\sum_S\mathfrak{f}_S^2/\bar{n}_S$, is the total time-independent Poisson contribution. The adiabatic IC is $P_\delta^{(\rm ad)}(y_0,k)\approx 36P_{\mathcal{R}}(k)\left[3+\ln\ml(0.15k/k_{\mathrm{eq}}\mr) - \ln\ml(4/y_0\mr)\right]^2$, with $k^3/(2\pi^2)P_\mathcal{R}(k)\approx 2\times 10^{-9}$ \cite{Planck:2018jri}.

The three different 
$\mathcal{T}_k^{(\mathrm{a}, \mathrm{b}, \mathrm{c})}$ in the above expressions are determined by the following Volterra equations:\footnote{$\mathcal{T}_k^{(\mathrm{a,b})}$ describe the evolution of initial bulk perturbations to the density and the velocity divergence, respectively, whereas $\mathcal{T}_k^{(\mathrm{c})}$ is related to the evolution of the Poisson fluctuations.}
\beq
\label{eq:Ty}
    \mathcal{T}^{(i)}_k(y,y') &= {\mathcal{T}}^{\mathrm{fs}\,(i)}_k(y,y')+\frac{3}{2}\int_{y'}^y \frac{\dl y''}{\sqrt{1+y''}}{\mathcal{T}}^{\mathrm{fs}\,(\mathrm{b})}_k(y,y'') {\mathcal{T}}^{(i)}_k(y'',y')\,\quad i=\mathrm{a,b,c}.
\eeq
Solving these Volterra equations requires a specification of the free-streaming kernels, ${\mathcal{T}}^{\mathrm{fs}\,(\mathrm{a,b,c})}_k$, which can be calculated based on initial velocity distributions, $f^S(\bv)$, of every species, which is assumed to not evolve beyond redshifting of velocities. We define $\bv$ as the velocity today, meaning that at scale factor $a$, the velocity would be $\bv/a$. Defining a species wide ``building block" kernel:
\beq
\mathcal{T}^{\mathrm{fs}\,S}_k(y,y')\equiv \int_\bv f^S(\bv) \exp\ml[-i\hat{\bv} \cdot \hat{\bk}\sqrt{2}\frac{k}{k_{\mathrm{eq}}}\frac{v}{a_{\mathrm{eq}}}
    \mathcal{F}(y,y')\mr],
\eeq
the free-streaming kernels are given by
\beq
    {\mathcal{T}}^{\mathrm{fs}\,(\mathrm{a})}_k(y,y') &=  \sum_s\mathfrak{f}_S {\mathcal{T}}^{\mathrm{fs}\,S}_k(y,y'),\quad
    {\mathcal{T}}^{\mathrm{fs}\,(\mathrm{b})}_k(y,y') = \mathcal{F}(y,y'){\mathcal{T}}^{\mathrm{fs}\,(\mathrm{a})}_k(y,y'),\\
    \quad
{\mathcal{T}}^{\mathrm{fs}\,(\mathrm{c})}_k(y,y') &= \frac{1}{P_\delta^{(\rm iso)}(y_0,k)}\sum_S\frac{\mathfrak{f}^2_S}{\bar{n}_S} \mathcal{T}^{\mathrm{fs}\,S}_k(y,y').
\eeq
Here, $\mathcal{F}(y,y') = \ln\ml[(y/y')(1+\sqrt{1+y'})^2 / (1+\sqrt{1+y})^2\mr]$ captures the functional dependence of the comoving distance traveled by a particle during the time interval between $y'$ and $y$. Note that the Volterra equations \eqref{eq:Ty} are the same as the single species case, only with more complicated ``initial'' functions ($\mathcal{T}^{\mathrm{fs}(i)}$) obtained from the weighted sums of the initial phase space distribution functions of all the species.

\subsection{Inter/Intra-species Power Spectra}\label{sec:crosspower}
The above result is for the power spectrum for the total density contrast. It is also possible to obtain more detailed information related to different species. The cross power spectrum of $\mathfrak{f}_S\delta_S$ and $\mathfrak{f}_{S'}\delta_{S'}$ is
\beq\label{eq:Pss}
P_\delta^{SS'}(k)
&=
\mathfrak{f}_S \frak{f}_{S'}P_\delta^{(\mathrm{ad})}(y_0,k)\mathcal{T}_k^{(\mathrm{ad})S}(y,y_0)\mathcal{T}_{k}^{(\mathrm{ad})S'}(y,y_0)+
\\
&\phantom{=}\frac{\frak{f}_S^2}{\bar{n}_S}\delta_{SS'}+
\frac{3}{2}\,\frak{f}_S \frak{f}_{S'}P_\delta^{(\mathrm{iso})}(y_0,k)\int_{y_0}^y \frac{\dl y'}{\sqrt{1+y'}} \left[
\mathcal{T}^{(\mathrm{b})S}_{k}(y,y')\mathcal{T}^{(\mathrm{c})S'}_{k}(y,y')
+(S\leftrightarrow S')\right].
\eeq
Note that by definition $P_{\delta}=\sum_S\sum_{S'}P_\delta^{SS'}$.
The Volterra equations that need to be solved now are coupled across species (note the summation in the last term below):
\beq
\label{eq:Tis}
    \mathcal{T}^{(i)S}_k(y,y') &= {\mathcal{T}}^{\mathrm{fs}\,(i)S}_k(y,y')+\frac{3}{2}\int_{y'}^y \frac{\dl y''}{\sqrt{1+y''}}{\mathcal{T}}^{\mathrm{fs}\,(\mathrm{b})S}_k(y,y'')\sum_{S'} \mathfrak{f}_{S'}{\mathcal{T}}^{(i)S'}_k(y'',y')\,,
\eeq
where $i=\mathrm{a,b,c}$. The adiabatic transfer function $\mathcal{T}^{(\mathrm{ad})S}_k$ for each species is still given by the second line of \eqref{eq:MainResultT}, with $(i)\rightarrow (i)S$ in the superscript. The free-streaming kernels ${\mathcal{T}}_k^{\mathrm{fs}\,(\mathrm{a})S}={\mathcal{T}}_k^{\mathrm{fs}\,S}$,${\mathcal{T}}_k^{\mathrm{fs}\,(\mathrm{b})S}=\mathcal{F}{\mathcal{T}}_k^{\mathrm{fs}\,S}$, and ${\mathcal{T}}_k^{\mathrm{fs}(\mathrm{c})S}=({\mathfrak{f}_S/\bar{n}_S}/{P       ^{({\rm iso})}_\delta}){\mathcal{T}}_k^{\mathrm{fs}\,S}$.

\section{Examples}
\label{sec:Examples}
When each component has an initial Maxwell-Boltzmann distribution with characteristic co-moving velocity dispersion $\sigma_{* S}$, we have
\beq
f^S(\bv)=(2\pi\sigma_{*S}^2)^{-3/2}e^{-\frac{v^2}{2\sigma_{*S}^2}},\quad 
{\mathcal{T}}^{\mathrm{fs}\,S}_k(y,y')=\exp[-\alpha_{k\,S}^2\mathcal{F}^2(y,y')/2],
\eeq
where we defined 
\beq\label{eq:seq_alpha}
\sigma_{{\rm eq}\,S}\equiv \sigma_{*S}/a_{\rm eq},\qquad \alpha_{k\,S}\equiv\sqrt{2}(k/k_{\rm eq})\sigma_{{\rm eq}\,S}.
\eeq
In the limit that $\sigma_{*S}\rightarrow 0$, we have $f^S(\bv)\rightarrow \delta_\mathrm{D}(\bv)$ and $\tilde{\mathcal{T}}^{\mathrm{fs}\,S}_k(y,y')\rightarrow 1$. \\ \\

\paragraph{Relevant Scales:} The Jeans scale and free-streaming scale for a species $S$ are given by
\beq
k_{\mathrm{J}\, S}(y)&=\frac{\sqrt{3y}}{2}\frac{k_{\rm eq}}{\sigma_{{\rm eq}\,S}}\approx 120\,\Mpc^{-1}\left(\frac{22\,{\rm km} \,s^{-1}}{\sigma_{{\rm eq}\,S}}\right)\sqrt{y},\\
k_{\mathrm{fs}\, S}(y)&=\frac{1}{\sqrt{2}\mathcal{F}(y,y_0)}\frac{k_{\rm eq}}{\sigma_{{\rm eq}\,S}}\approx 15\,\Mpc^{-1}\frac{\mathcal{F}(1,10^{-3})}{\mathcal{F}(y,y_0)}\left(\frac{22\,{\rm km}\, s^{-1}}{\sigma_{{\rm eq}\,S}}\right).
\eeq
These are the key scales determining the PS time evolution. By substituting these special comoving $k$ values in the definition for $\alpha_{k\,S}$ in equation~\eqref{eq:seq_alpha}, we have:
\beq\label{eq:alpha_scales}
\alpha_{\mathrm{J}\, S}(y)=\sqrt{3y/2}\,,\qquad \alpha_{\mathrm{fs}\, S}(y)=\mathcal{F}^{-1}(y,y_0).
\eeq
It is useful to note that $y=1$ at equality and that $\mathcal{F}^{-1}(1,10^{-3})=0.15$ and is essentially frozen at this value as we increase $y$. We plot our results in terms of $\alpha_{k\,S}$, which makes the results independent of particular choices of $\sigma_{*S}$ (this independence is exact for the isocurvature contribution but not for the adiabatic contribution). 

\begin{figure}
    \centering
    \includegraphics[width=1\linewidth]{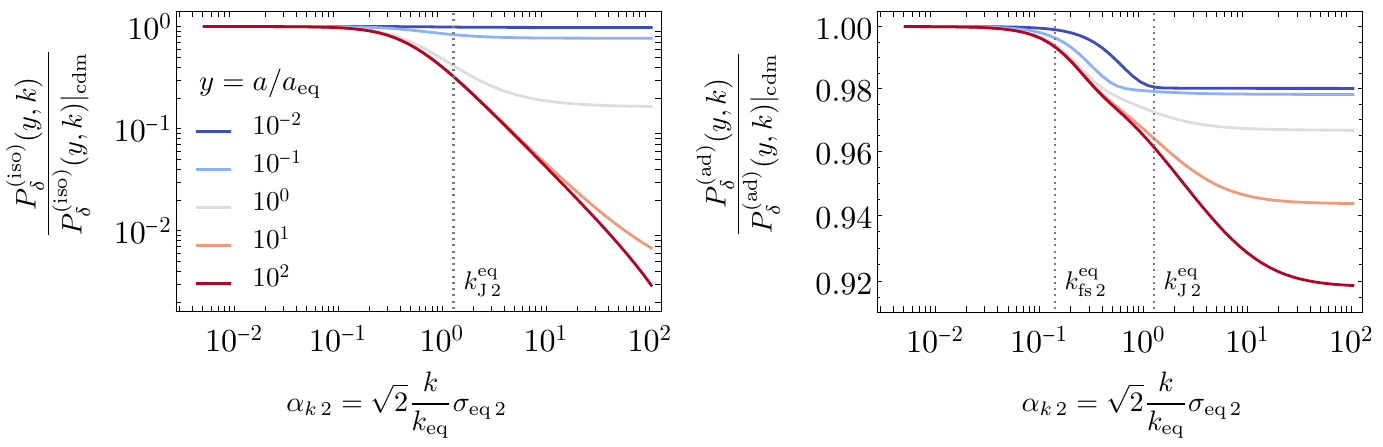}

    \caption{The isocurvature (left) and adiabatic (right) growth for 2-component dark matter compared to single-component CDM. The 2-component DM consists of a dominant CDM component without significant Poisson fluctuations (component 1) and a 1\% component that is warm and has significant Poisson fluctuations (component 2). For the isocurvature part of the power spectrum, the suppression due to the warm component begins at the Jeans scale at equality, corresponding to $\alpha_{k\, 2}\simeq \sqrt{3/2}$ (see equation~\ref{eq:alpha_scales}). For $y\gg 1$, suppression for larger $\alpha_{k\,2}$ scales as $ (4/9)\alpha_{k\,2}^{-1}$.
    For the adiabatic part, the suppression begins at the free-streaming scale $\alpha_{\mathrm{fs}\,2}(y)$. For $y\gg1$ the suppression begins around $\alpha_{k\, 2}\simeq 0.15$ (see equation~\ref{eq:alpha_scales}) and plateaus at the current Jeans scale $\alpha_{k\, 2}\simeq \sqrt{3y/2}$ with a plateau depth of $\approx (2/5)\mathfrak{f}_2(8+ 3\ln y)$. To convert the horizontal axis to wave number, use $k\approx 10^2\,\Mpc^{-1}\left({22\,{\rm km}\,{\rm s}^{-1}}/{\sigma_{{\rm eq}\,S}}\right)\alpha_{k\,S}.$}
    \label{fig:Case1}
\end{figure}

Let us restrict ourselves to two components for simplicity. We will make some further simplifying approximations for analytic tractability, but the Volterra equations can of course always be solved numerically without making these approximations. \ma{The plots of power spectra shown in this section are based on solving the Volterra equations using the algorithm and code provided in Appendix \ref{App:Num-PS}.}

\subsection{Case 1: Dominant cold adiabatic + subdominant warm Poisson}
We take component $1$ to be cold with $f^1(\bv)=\delta_D(\bv)$, whereas component $2$ has a Maxwell-Boltzmann distribution with characteristic velocity $\sigma_{2*}$. We also take $\bar{n}_1\rightarrow \infty$, and assume $\mathfrak{f}_2\ll 1$ for analytic tractability. Then $P_{\delta}^{(\rm iso)}(y_0,k)=\mathfrak{f}_2^2/\bar{n}_2$, ${\mathcal{T}}^{\mathrm{fs}\,(\mathrm{a})}_k=\mathfrak{f}_1+\mathfrak{f}_2e^{-\alpha_{k\,2}^2\mathcal{F}^2/2},\mathcal{T}^{\mathrm{fs}\,(\mathrm{b})}_k=\mathcal{F}(\mathfrak{f}_1+\mathfrak{f}_2e^{-\alpha_{k\,2}^2\mathcal{F}^2/2}),$ and ${\mathcal{T}}^{\mathrm{fs}\,(\mathrm{c})}_k=e^{-\alpha_{k\,2}^2\mathcal{F}^2/2}$. 

At zeroth order in $\mathfrak{f}_2$, the $\mathrm{a,b}$ transfer functions are given by
\beq
\mT{b}{k}(y,y')&=3 \left(1+\frac{3}{2} y\right) \left(1+\frac{3}{2} y'\right) \left[\frac{\sqrt{1+y}}{1+\frac{3}{2} y}-\frac{\sqrt{1+y'}}{1+\frac{3}{2} y'}-\frac{1}{3}\ln\left(\frac{x}{x'}\right)\right],\\
\mT{a}{k}(y,y')&=\left(1+\frac{3}{2} y\right) \left(1+3y'\right)-\frac{3}{2} y' \sqrt{1+y'} \left[3
   \sqrt{1+y}-\left(1+\frac{3}{2} y\right) \ln \left(\frac{x}{x'}\right)\right],
\eeq
where $x\equiv\frac{\sqrt{1+y}+1}{\sqrt{1+y}-1}$ and $x'\equiv\frac{\sqrt{1+y'}+1}{\sqrt{1+y'}-1}$. At leading non-trivial order, we can also find\footnote{Notice that we switched the roles of $\mathcal{T}^{(\mathrm{b})}$ and $\mathcal{T}^{(\mathrm{c})}$ inside the integrand in the second line. To see that this is allowed, see Appendix A of \cite{Amin:2025dtd}, and footnote 6 of \cite{Amin:2025sla}.}
\beq
\mathcal{T}^{(\mathrm{c})}_{k}(y,y')&= e^{-\frac{\alpha_{k\,2}^2\mathcal{F}^2(y,y')}{2}}+\frac{3}{2}\int_{y'}^y \frac{\dl y''}{\sqrt{1+y''}}\mathcal{T}^{(\mathrm{b})}_{k}(y,y'')e^{-\frac{\alpha_{k\,2}^2\mathcal{F}^2(y'',y')}{2}}.
\eeq
These solutions form the basis for a perturbative construction of the adiabatic and isocurvature spectra at leading non-trivial order in $\mathfrak{f}_2$.

\paragraph{Isocurvature:} 
For $\alpha_{k\,2}\ll 1$ and $\alpha_{k\,2}\gg 1$, we can find simple formulae:
\begin{equation}
 P^{(\rm iso)}_{\delta}(y, k)
   \approx P^{(\rm iso)}_\delta(y_0,k)
\begin{cases}
\left(1+\frac{3}{2} y\right)^2 &\alpha_{k\,2}\ll 1,\\
   1+3y/\alpha_{k\,2}^{2}+y^2/\alpha_{k\,2}&\alpha_{k\,2}\gg 1,y\ll 1,\\
   1+y^2/\alpha_{k\,2}&\alpha_{k\,2}\gg 1,y\gg 1.
\end{cases}
\end{equation}
The $\alpha_{k\,2}\gg 1$ cases are fits, and work well ($\sim 10\%$) for $10<\alpha_{k\,2} < 100$. The last line shows that the growth at late times is $y^2$, however it has a shallow scale dependence of $k^{-1}$. Compare this to the case of $k^{-4}$ expected for a single warm component case in the same regime \cite{Amin:2025sla}. The presence of the dominant cold component ameliorates the Jeans suppression from the warmth in the subdominant component. 

To quantitatively understand this, we can evaluate $P^{11}(y,k)$ (from \eqref{eq:Pss}) to quadratic order in $\mathfrak{f}_2$. After $P^{11}(y,k)$ grows to be of the same order as the Poisson-noise floor, $P^{22}(y_0,k)$, the influence of component 2 can be neglected, 
and we have CDM-like growth in component 1 for that $k$ mode. Upon evaluation,\footnote{Again, this can be done without solving Volterra equations.} one can show that the state $P^{11}(y,k)\sim P^{22}(y_0,k)$ is reached when $y=y_k\sim \sqrt{\alpha_{k\,2}}$ for $\alpha_{k\,2}>1 $. Hence, the growth of the power spectrum at large $k$ and late times is $P^{11}_\delta(y,k)\sim (\mathfrak{f}_2^2/\bar{n}_2)(y/y_k)^2\propto 1/k$. 

\sd{
Intuitively, the $k^{-1}$ scaling can be understood from the following physical argument. Particles of species 2 with velocities $\sigma_2$ cross a comoving length scale $k^{-1}$ in time $\sim a/(k\sigma_2)$. Consequently, on the scale $k^{-1}$, density perturbations due to Poissonian particle noise only remain coherent for a time scale $\tau_k = a/(k\sigma_2)$.
Meanwhile, the time scale for perturbation growth is the Hubble time; perturbation growth is effectively sourced by the average density over that time.
For particle noise, the density contrast can be viewed as taking $N_k=H^{-1}\tau_k^{-1}$ independent values over that time scale, all drawn from the same Poisson distribution.
Consequently, the mean squared value of the Hubble-time-averaged density contrast is $\sim 1/N_k$ times the mean squared value of the instantaneous density contrast. So the power spectrum sourced by the particle noise is scaled by $\sim 1/N_k = a H / (k \sigma_2) \sim k_{\mathrm{J}\, 2}/k$.
}

\paragraph{Adiabatic:} For the adiabatic part, we must still solve for $\mathcal{T}_k^{(\rm a,b)}$ to linear order in $\mathfrak{f}_2$. Again, by doing a perturbative calculation, we do not have to solve any Volterra equations. There is a gentle suppression of the power spectra at large $\alpha_{k\,2}$ of the form:
\beq
P^{(\rm ad)}_\delta(y,k)\approx P^{(\rm ad)}_\delta(y,k)|_{\rm cdm}
\begin{cases}
1 &\alpha_{k\,2}\ll 1,\\
1 -\frac{2}{5}\mathfrak{f}_2(7.8+3\ln y)&\alpha_{k\,2}\gg 1,y\gg 1,\\
\end{cases}
\eeq
where $\mT{ad}{k}$ refers to the adiabatic transfer function for CDM (at zeroth order in $\mathfrak{f}_2$). 
We note the presence of a logarithmically growing mode similar to that identified by \cite{Celik:2025wkt}.

\begin{figure}
    \centering
    \includegraphics[width=1\linewidth]{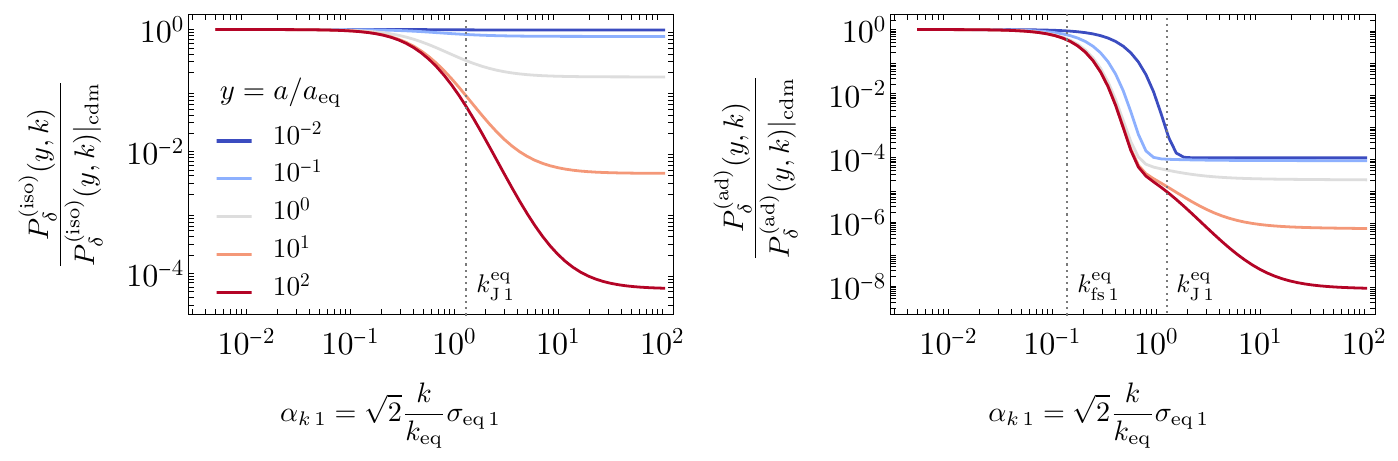}

    \caption{The isocurvature (left) and adiabatic (right) growth of PS compared to CDM for 2-component dark matter, with 1\% cold with significant Poisson fluctuations (rest warm dark matter). For the isocurvature part, the suppression from unity due to the warm component begins at the Jeans scale for the warm component at equality $\alpha_{\mathrm{J}\,1}(y=1)=\sqrt{3/2}$ and plateaus $\alpha_{\mathrm{J}\,1}(y)=\sqrt{3y/2}\gg 1$. The depth of the suppression is  $\approx (4y^{-2}/9)[1+6 \mathfrak{f}_2\ln(y/4)]$ at large $y$. For the adiabatic part, the suppression begins at the free-streaming scale $\alpha_{\mathrm{fs}\,1}=\mathcal{F}^{-1}(y,y_0)$. For $y\gg1$ the suppression plateaus at the current Jeans scale $\alpha_{\mathrm{J}\,1}(y)$. The height of this suppressed part $\sim \mathfrak{f}_2^2 4y^{-2}/9$. To convert the horizontal axis to wave number, use
$k\approx 10^2\,\Mpc^{-1}\left({22\,{\rm km} s^{-1}}/{\sigma_{{\rm eq}\,1}}\right)\alpha_{k\,1}.$}
    \label{fig:Case2}
\end{figure}
\subsection{Case 2: Dominant warm adiabatic + subdominant cold Poisson}
We now consider an example where the dominant component 1 is warm while the subdominant (1\%) component 2 is cold. We still allow a significant Poisson contribution to component 2 only. In this case, the $\mT{a,b}{k}$ at zeroth order in $\mathfrak{f}_2$ have to be evaluated numerically. They include the free-streaming suppression of the adiabatic spectrum common to warm dark matter.

\paragraph{Isocurvature:} The total power spectrum is suppressed beyond the Jeans scale of the dominant component ($\alpha_{k\,1}\gtrsim 1$). The subdominant cold component has a white noise contribution for $\alpha_{k\,1}\gg 1$. For $\mathfrak{f}_2\ll 1$, this cold component grows in a scale invariant fashion. This growth follows $P^{(\rm iso)}_\delta (y,k)\approx P_\delta^{(\rm iso)}(y_0,k)\left[1+12\mathfrak{f}_2\ln \left(\frac{1+\sqrt{1+y}}{1+\sqrt{1+y_0}}\right)\right]$, 
where we assumed $y_0\ll 1$.

\paragraph{Adiabatic:} On scales where $\alpha_{k\,1}\lesssim 1$, $\mT{ad}{k}$ is similar to the case of a single warm component.
For $\alpha_{k\,1}\gtrsim 1$, the total power spectrum behavior becomes $\propto y^2/\alpha_{k\,1}^2$. It becomes independent of $\alpha_{k\,1}$ for $\alpha_{k\,1}\gg 1$. Carrying out a perturbative calculation in $\mathfrak{f}_2$ using \eqref{eq:Pss} and \eqref{eq:Tis}, we get $P_{\delta}^{(\rm ad)}(y,k)\approx \mathfrak{f}_2^2 P^{(\rm ad)}_\delta(y_0,k)\left[1+\frac{1}{2}\frac{\dl\ln P_\delta(y_0,k)}{\dl \ln y} \mathcal{F}(y,y_0)\right]^2$,
for $y\gg 1$.

\begin{figure}
    \centering
    \includegraphics[width=0.95\linewidth]{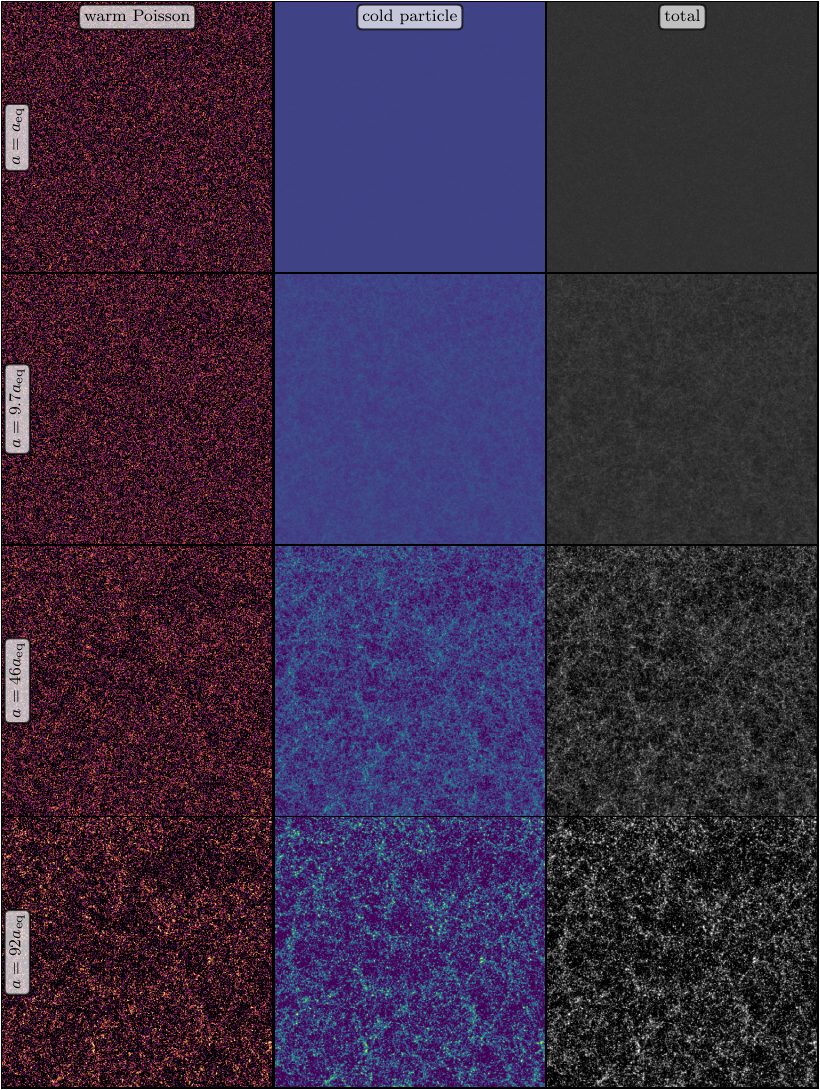}

    \caption{\sd{Growth of structure in a cosmological simulation of a two-component dark matter model. The subdominant component ($\mathfrak{f}_2=0.03$) is warm ($\sigma_{{\rm eq}\,2}\approx 65\,{\rm km}\, {\rm s}^{-1}$) and has massive particles ($m_2\approx 2\times 10^4M_{\odot}$), with correspondingly significant Poisson fluctuations, while the dominant component is usual CDM. We show the projected density field of each component, along with their sum. The dominant cold component is initialized without density perturbations, but the subdominant component seeds structure (above its Jeans length) in the dominant component. At the level of the power spectrum, the co-evolution of these two components is captured well by our analytic framework.}
    }
    \label{fig:figure-fields}
\end{figure}
\begin{figure}
    \centering
    \includegraphics[width=6.1 in]{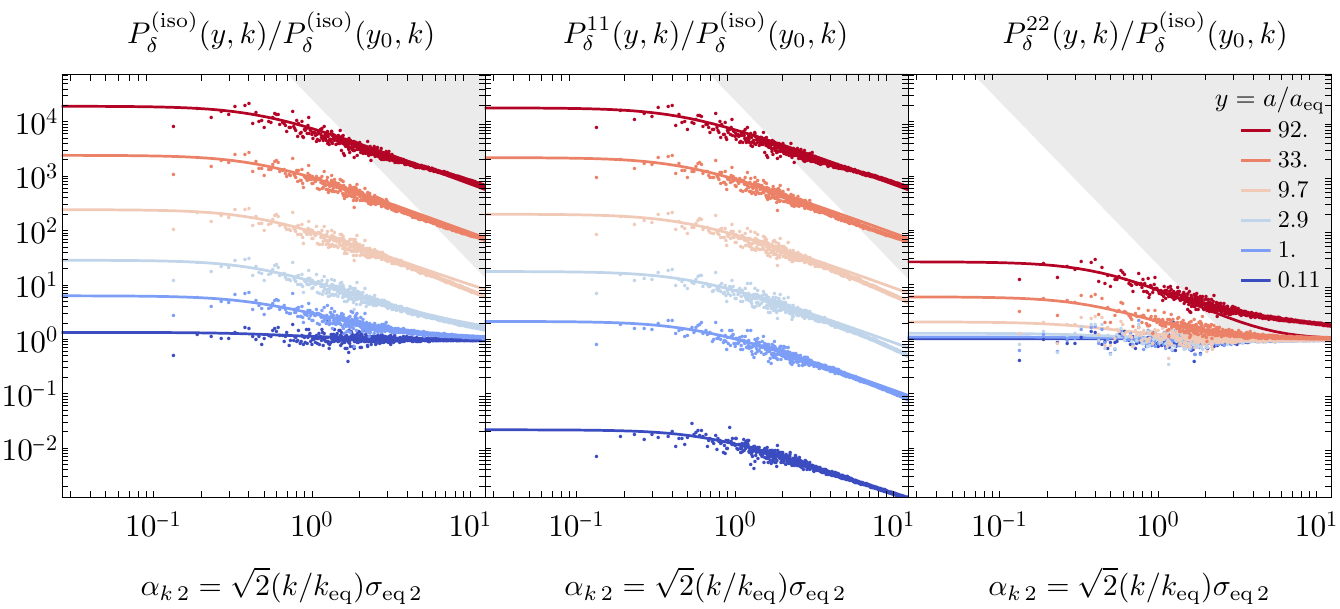}
    \caption{The scale-dependent isocurvature growth in a two species dark matter model where the first component is cold and without significant Poisson fluctuations, whereas the subdominant second component ($\mathfrak{f}_2=0.03$) is warm ($\sigma_{{\rm eq}\,2}\approx 65\,{\rm km}\, {\rm s}^{-1}$) and has large Poisson fluctuations ($\bar{n}_2\approx 4.5\times 10^4\,\Mpc^{-3}$). The solid curves are from our analytic calculations, while the dots are based on an $N$-body simulation. Along with the total power spectrum \ma{(left-most panel)}, we also show the \ma{intra-species} power spectra. \ma{The gray shaded region indicates nonlinear evolution, $k^3P^{(i)}_\delta(y,k)/2\pi^2>1$, with $i={\rm iso},11$ and $22$.} The analytics and $N$-body simulation results agree in the linear regime (and the total matter and species 1 power spectra surprisingly even agree in the nonlinear regime). \ma{Some deviations between the $N$-body results and linear power spectra predictions  appear at $k^3P^{22}_\delta(y,k)/2\pi^2>1$, even when the dominant component is in the linear regime.}}
    \label{fig:NumCompare}
\end{figure}

\subsection{\sd{Validation with $N$-body Simulations}}
\label{sec:N-body}
\sd{To test our results, we simulated the formation of structure in a 2-component dark matter scenario using $N$-body simulations.} A subdominant second component ($\mathfrak{f}_2= 0.03$) is warm and has significant Poisson fluctuations in density, while the dominant component is cold and without significant Poisson fluctuations. The initial conditions for the dominant cold component are set by starting with particles at rest on a grid. The subdominant component is initialized by drawing particle positions at random and giving them random velocities from a Maxwell-Boltzmann distribution. We used $N_1\approx 10^8$ particles and $N_2\approx 10^5$ particles ($m_2\approx 2\times 10^4M_\odot,\sigma_{{\rm eq}\,2}\approx 22\,{\rm km}\, s^{-1}$). The co-moving box size was $\sim \Mpc$. The simulation does not capture the adiabatic perturbations (which were included in \cite{Amin:2025dtd}), but \sd{they would be subdominant to the Poisson fluctuations on almost all scales small enough to be represented in the simulation volume.}

\sd{We executed the simulation using the \textsc{Gadget} 4 cosmological simulation code \cite{Springel:2020plp}. Figure~\ref{fig:figure-fields} shows the projected density in each species, as well as the projected total density, over time. The subdominant species with Poisson fluctuations can be seen to source structure in the initially unperturbed dominant species. Meanwhile, the points in Fig.~\ref{fig:NumCompare} show the evolution of the power spectrum and its different components in the simulation. This figure shows how}
once the density perturbations in the dominant species exceed those in the subdominant species at a given scale, the dominant species evolves essentially like CDM. The shape of the total isocurvature power spectrum $P^{(\rm iso)}_\delta$ reflects this effectively delayed growth. Along with the total power spectrum, we also show the intra-species ($P^{11}_\delta$ and $P^{22}_\delta$) 
power spectra evaluated from the simulation.

\sd{For comparison, the solid curves in Fig.~\ref{fig:NumCompare} show the power spectra obtained through our analytic formalism by solving the Volterra equations \eqref{eq:Tis}. Overall, the analytic and $N$-body results agree well as long as the perturbations remain in the linear regime.
At late times, perturbations in the dominant cold species start to become nonlinear, corresponding to $k^3P^{11}_\delta/2\pi^2\gtrsim 1$. Modest deviations from the analytic results start to appear in this regime, especially in $P_{22}$.\footnote{\sd{For nonlinear perturbations to the dominant species 1, $P^{11}_\delta$ itself remains remarkably close to the analytic prediction. However,} this is likely a coincidence. The same $P^{(\rm iso)}_\delta\propto k^{-1}$ nonlinear matter power spectrum arose in the simulations of \cite{Amin:2025dtd} for a different scenario, where the power spectrum predicted by the formalism was different.}
Meanwhile, subtle deviations from the analytic results are visible even at early times for sufficiently high $k$ (specifically $\alpha_{k\,2}\gtrsim 5$). These deviations are likely due to nonlinearity in perturbations to the subdominant component 2 (corresponding to $k^3P^{22}_\delta/2\pi^2\sim \mathfrak{f}_2^2$), which is present from the outset.\footnote{\sd{For context, we note that length scales on which perturbations to component 2 are initially nonlinear roughly correspond to volumes inside which fewer than one component-2 particle is expected.}}
We ran a similar simulation with a lower velocity dispersion $\sigma_{{\rm eq},2}$ (and hence a higher Jeans wavenumber), and we verified that these deviations remained aligned with the scale at which $P^{22}_\delta$ approaches nonlinearity.
}

\begin{figure}
    \centering
    \includegraphics[width=1\linewidth]{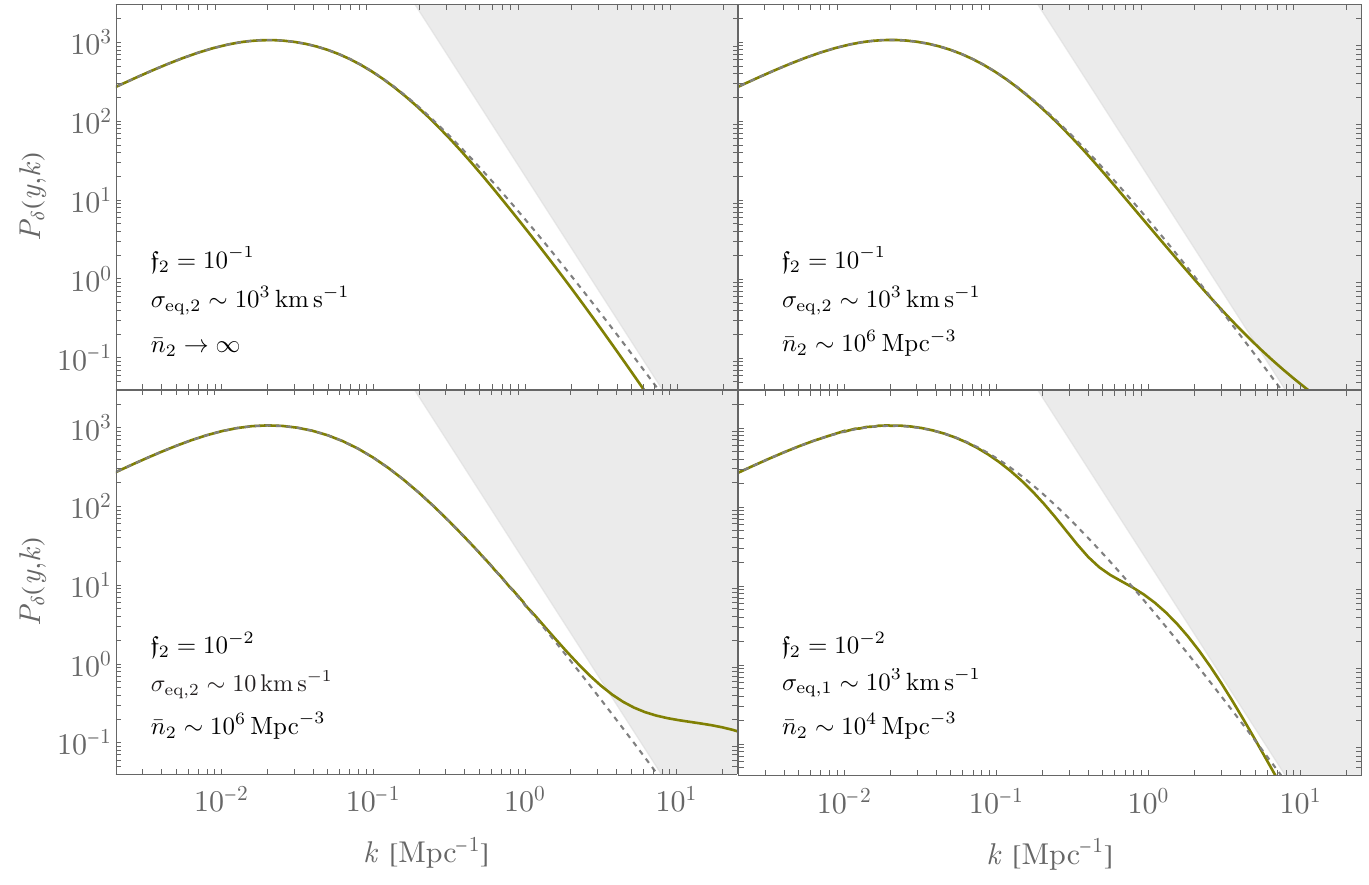}
    \caption{\ma{Examples of linear dark matter power spectra at $y= 300$ ($z\approx 10$) in dark matter models with two components. The mass fractions, velocity dispersion, and Poisson noise levels are varied. We assume that the second component is always subdominant and has the Poisson noise, but either component may be warm. We use the (approximately) parameter independent transfer functions in Fig.~\ref{fig:Case1} and \ref{fig:Case2} to construct the above examples by appropriate scalings. The wavenumber of departure from CDM power spectra, and the amplitude and shape of the departure, can be controlled by choosing the mass fractions $\mathfrak{f}_S$, the velocity dispersions at equality $\sigma_{\rm eq,S}$ and the number densities $\bar{n}_{S}$. The gray region indicates where evolution is nonlinear.}}
    \label{fig:Examples}
\end{figure}

\subsection{Observational context}

Galaxy surveys can typically probe deviation of the power spectrum at $k_{\rm obs}\sim 0.1$-$1\,\Mpc^{-1}$ at the level of a few tens of percent \cite{DES:2021wwk,DESI:2024mwx}. The Lyman $\alpha$ forest and high-redshift luminosity function observations can reach smaller scales $k_{\rm obs}\gtrsim 10\,\Mpc^{-1}$ with order-unity deviations allowed \cite{Chabanier:2019eai,Sabti:2021unj}. Similarly, different observations will have their own $k_{\rm obs}$ range and tolerance for deviations from $\Lambda$CDM expectations. Heuristically, the parameters $(\mathfrak{f}_S, \bar{n}_S, \sigma_{{\rm eq}\,S})$ of interest observationally are those for which $k_{{\rm fs}\, S}^{\rm eq}\sim k_{\rm obs}$  and $\mathfrak{f}_S^2/\bar{n}_S\sim P^{(\Lambda {\rm CDM})}_\delta(y,k_{\rm obs})$. However, we note that there is a broader range of parameters whose combination leads to non-trivial effects in the power spectrum on observable scales. 
For example, it is possible for the suppression of power from warm dark matter and the enhancement of power from Poisson noise to partially compensate, leaving only a weak imprint on the power spectrum (e.g.~\cite{Tadepalli:2025gzf}).
In Fig.~\ref{fig:Examples}, we show matter power spectra for a range of different 2-component scenarios with different levels of warmth, Poisson noise, and mass fractions.

\ma{As we noted in Section~\ref{sec:Main}, our results apply directly only when perturbations are small. In Fig.~\ref{fig:Examples}, we mark the regime of large perturbations, $k^3P_\delta(k,z)/2\pi^2\gtrsim 1$.\footnote{\ma{A more detailed assessment of the domain of validity can be done by determining whether each component's evolution is nonlinear. In Sec.~\ref{sec:N-body}, when comparing $N$-body simulations and our linear power spectra predictions, subtle deviations begin to appear at scales where the subdominant component becomes nonlinear (see Fig.~\ref{fig:NumCompare}).}}
There, while the perturbative $P_\delta(k,z)$ that we plot is not necessarily a direct measure of clustering, it can still be used to predict halo mass functions as done in \cite{Amin:2025dtd}. This is similar to how linear power spectra are used in Press-Schechter or excursion-set formulations.
}

\section{Summary \& Conclusions}
\label{sec:Summary}
We have provided a general framework to calculate scale-dependent power spectrum evolution for multicomponent dark matter during matter and radiation domination. The components can be cold or warm and can have significant Poisson fluctuations associated with their discrete nature.
Any combination of these features can be present in each component, and the framework requires no restrictions on the number of components or their mass fraction. We do not assume that our components are perfect fluids, and include effects of velocity dispersion as well as discreteness of the effective constituent ``particles".

We are able to evaluate the total power spectrum, as well as cross and self power spectra for density perturbations in each component. Along with the total spectrum, cross spectra provide insights on how one component affects the other. We have provided a numerical algorithm (see Appendix \ref{App:Num-PS}, and publicly available code at \url{https://github.com/mustafaaamin/warm-and-random.git}) to evaluate these spectra. Typically, total power spectra calculations can be done in seconds. 

We provided explicit examples of power spectrum calculations. In particular, we considered a warm or cold subdominant component with significant Poisson noise. Primordial black holes, solitons, and miniclusters, as well as interference granules of wave dark matter produced after inflation, provide motivating examples where our framework can be applied. It can also be applied to cases such as with neutrinos (approximately) and with fractional or dominant amounts of warm dark matter. 

Our framework is restricted to subhorizon, non-relativistic dynamics, although this is not a severe restriction for most models of dark matter that are still viable. Also, like standard cosmological perturbation theory, the power spectrum calculation in this work loses control once the growth of perturbations become nonlinear. We also note that while the framework applies to interference granules, solitons, and such on scales larger than their size and separation, finite size effects (e.g. at the de Broglie scale) have not been accounted for in the present work. Such effects were included in \cite{Amin:2025sla} for a single species. A companion paper \cite{Amin:2025nxm} includes these wave effects for the multi-species case.

We have ignored all non-gravitational effects. Our framework can be extended to include additional non-gravitational interactions between dark matter particles \cite{Spergel:1999mh,Bottaro:2024pcb,Bottaro:2023wkd,Graham:2025fdt,Costa:2025kwt}. We leave this extension for future work. We also note that incorporating our multi-species DM evolution into existing cosmological perturbation solvers \cite{2022ascl.soft03026G, 2011JCAP...09..032L,Chudaykin:2020aoj} that include baryonic effects would be useful to understand the effects on BAO and CMB scales. 

\section*{Acknowledgements}
MA is supported by a DOE award DE-SC0010103. Simulations for this work were carried out on the OBS HPC computing cluster at the Observatories of the Carnegie Institution for Science. MA acknowledges helpful conversations with Marco Simonovic, Elisabeth Krause and Pranjal R.S. regarding observational feasibility, Ennio Salvioni, Francesco Verdiani and Fabian Schmidt about their work on EFT of mixed dark matter, and thanks the Galileo Galilei Institute for their hospitality.

\bibliographystyle{JHEP}
\bibliography{main}

@article{Costa:2025kwt,
    author = "Costa, Marco and Creque-Sarbinowski, Cyril and Simon, Olivier and Weiner, Zachary J.",
    title = "{Dark forces suppress structure growth}",
    eprint = "2510.00098",
    archivePrefix = "arXiv",
    primaryClass = "astro-ph.CO",
    month = "9",
    year = "2025"
}

@book{mayer1948statistical,
  title={Statistical Mechanics},
  author={Mayer, J.E. and Mayer, M.G.},
  url={https://books.google.com/books?id=IrF-0AEACAAJ},
  year={1948},
  publisher={J. Wiley \& sons, Incorporated}
}

@BOOK{Binney:1987,
       author = {{Binney}, James and {Tremaine}, Scott},
        title = "{Galactic dynamics}",
         year = 1987,
       adsurl = {https://ui.adsabs.harvard.edu/abs/1987gady.book.....B}
}

@article{Kolb:1993zz,
	archiveprefix = {arXiv},
	author = {Kolb, Edward W. and Tkachev, Igor I.},
	doi = {10.1103/PhysRevLett.71.3051},
	eprint = {hep-ph/9303313},
	journal = {Phys. Rev. Lett.},
	pages = {3051--3054},
	reportnumber = {FERMILAB-PUB-93-066-A},
	title = {{Axion miniclusters and Bose stars}},
	volume = {71},
	year = {1993}}

@article{Arvanitaki:2009fg,
	archiveprefix = {arXiv},
	author = {Arvanitaki, Asimina and Dimopoulos, Savas and Dubovsky, Sergei and Kaloper, Nemanja and March-Russell, John},
	doi = {10.1103/PhysRevD.81.123530},
	eprint = {0905.4720},
	journal = {Phys. Rev. D},
	pages = {123530},
	primaryclass = {hep-th},
	title = {{String Axiverse}},
	volume = {81},
	year = {2010}}

@ARTICLE{2011JCAP...09..032L,
       author = {{Lesgourgues}, Julien and {Tram}, Thomas},
        title = "{The Cosmic Linear Anisotropy Solving System (CLASS) IV: efficient implementation of non-cold relics}",
      journal = {\jcap},
         year = 2011,
        month = sep,
       volume = {2011},
       number = {9},
          eid = {032},
        pages = {032},
          doi = {10.1088/1475-7516/2011/09/032},
archivePrefix = {arXiv},
       eprint = {1104.2935},
 primaryClass = {astro-ph.CO},
       adsurl = {https://ui.adsabs.harvard.edu/abs/2011JCAP...09..032L}
}

@article{Schive:2014dra,
	archiveprefix = {arXiv},
	author = {Schive, Hsi-Yu and Chiueh, Tzihong and Broadhurst, Tom},
	doi = {10.1038/nphys2996},
	eprint = {1406.6586},
	journal = {Nature Phys.},
	pages = {496-499},
	primaryclass = {astro-ph.GA},
	slaccitation = {%%CITATION = ARXIV:1406.6586;%%},
	title = {{Cosmic Structure as the Quantum Interference of a Coherent Dark Wave}},
	volume = {10},
	year = {2014}}

@article{Hui:2016ltb,
	archiveprefix = {arXiv},
	author = {Hui, Lam and Ostriker, Jeremiah P. and Tremaine, Scott and Witten, Edward},
	doi = {10.1103/PhysRevD.95.043541},
	eprint = {1610.08297},
	journal = {Phys. Rev. D},
	number = {4},
	pages = {043541},
	primaryclass = {astro-ph.CO},
	title = {{Ultralight scalars as cosmological dark matter}},
	volume = {95},
	year = {2017}}

@article{2022ascl.soft03026G,
       author = {{Grin}, Daniel and {Marsh}, David J.~E. and {Hlozek}, Renee},
        title = "{axionCAMB: Modification of the CAMB Boltzmann code}",
 howpublished = {Astrophysics Source Code Library, record ascl:2203.026},
         year = 2022,
        month = mar,
          eid = {ascl:2203.026},
       adsurl = {https://ui.adsabs.harvard.edu/abs/2022ascl.soft03026G}
}

@article{Spergel:1999mh,
    author = "Spergel, David N. and Steinhardt, Paul J.",
    title = "{Observational evidence for selfinteracting cold dark matter}",
    eprint = "astro-ph/9909386",
    archivePrefix = "arXiv",
    doi = "10.1103/PhysRevLett.84.3760",
    journal = "Phys. Rev. Lett.",
    volume = "84",
    pages = "3760--3763",
    year = "2000"
}

@article{Eberhardt:2025caq,
    author = "Eberhardt, Andrew and Ferreira, Elisa G. M.",
    title = "{Ultralight fuzzy dark matter review}",
    eprint = "2507.00705",
    archivePrefix = "arXiv",
    primaryClass = "astro-ph.CO",
    month = "7",
    year = "2025"
}

@article{Cirelli:2024ssz,
    author = "Cirelli, Marco and Strumia, Alessandro and Zupan, Jure",
    title = "{Dark Matter}",
    eprint = "2406.01705",
    archivePrefix = "arXiv",
    primaryClass = "hep-ph",
    month = "6",
    year = "2024"
}

@article{Amin:2022nlh,
    author = "Amin, Mustafa A. and Mirbabayi, Mehrdad",
    title = "{A Lower Bound on Dark Matter Mass}",
    eprint = "2211.09775",
    archivePrefix = "arXiv",
    primaryClass = "hep-ph",
    doi = "10.1103/PhysRevLett.132.221004",
    journal = "Phys. Rev. Lett.",
    volume = "132",
    number = "22",
    pages = "221004",
    year = "2024"
}

@article{Planck:2018jri,
    author = "Akrami, Y. and others",
    collaboration = "Planck",
    title = "{Planck 2018 results. X. Constraints on inflation}",
    eprint = "1807.06211",
    archivePrefix = "arXiv",
    primaryClass = "astro-ph.CO",
    doi = "10.1051/0004-6361/201833887",
    journal = "Astron. Astrophys.",
    volume = "641",
    pages = "A10",
    year = "2020"
}

@article{Drlica-Wagner:2022lbd,
    author = "Drlica-Wagner, Alex and others",
    title = "{Report of the Topical Group on Cosmic Probes of Dark Matter for Snowmass 2021}",
    eprint = "2209.08215",
    archivePrefix = "arXiv",
    primaryClass = "hep-ph",
    reportNumber = "FERMILAB-FN-1211-PPD",
    month = "9",
    year = "2022"
}

@article{Nadler:2024ims,
    author = "Nadler, Ethan O. and Gluscevic, Vera and Driskell, Trey and Wechsler, Risa H. and Moustakas, Leonidas A. and Benson, Andrew and Mao, Yao-Yuan",
    title = "{Forecasts for Galaxy Formation and Dark Matter Constraints from Dwarf Galaxy Surveys}",
    eprint = "2401.10318",
    archivePrefix = "arXiv",
    primaryClass = "astro-ph.GA",
    doi = "10.3847/1538-4357/ad3bb1",
    journal = "Astrophys. J.",
    volume = "967",
    number = "1",
    pages = "61",
    year = "2024"
}

@article{Gorghetto:2024vnp,
    author = "Gorghetto, Marco and Hardy, Edward and Villadoro, Giovanni",
    title = "{More axion stars from strings}",
    eprint = "2405.19389",
    archivePrefix = "arXiv",
    primaryClass = "hep-ph",
    reportNumber = "DESY-24-075",
    doi = "10.1007/JHEP08(2024)126",
    journal = "JHEP",
    volume = "08",
    pages = "126",
    year = "2024"
}

@article{Eggemeier:2019khm,
    author = "Eggemeier, Benedikt and Redondo, Javier and Dolag, Klaus and Niemeyer, Jens C. and Vaquero, Alejandro",
    title = "{First Simulations of Axion Minicluster Halos}",
    eprint = "1911.09417",
    archivePrefix = "arXiv",
    primaryClass = "astro-ph.CO",
    doi = "10.1103/PhysRevLett.125.041301",
    journal = "Phys. Rev. Lett.",
    volume = "125",
    number = "4",
    pages = "041301",
    year = "2020"
}

@article{Zhang:2024bjo,
    author = "Zhang, Hong-Yi",
    title = "{Unified view of scalar and vector dark matter solitons}",
    eprint = "2406.05031",
    archivePrefix = "arXiv",
    primaryClass = "hep-ph",
    month = "6",
    year = "2024"
}

@article{Lewis:2002nc,
    author = "Lewis, Antony and Challinor, Anthony",
    title = "{Evolution of cosmological dark matter perturbations}",
    eprint = "astro-ph/0203507",
    archivePrefix = "arXiv",
    doi = "10.1103/PhysRevD.66.023531",
    journal = "Phys. Rev. D",
    volume = "66",
    pages = "023531",
    year = "2002"
}

@article{Ji:2024ott,
    author = "Ji, Lingyuan and Dai, Liang",
    title = "{Effects of Subhalos on Interpreting Highly Magnified Sources Near Lensing Caustics}",
    eprint = "2407.09594",
    archivePrefix = "arXiv",
    primaryClass = "astro-ph.GA",
    month = "7",
    year = "2024"
}

@article{Delos:2021ouc,
    author = "Delos, M. Sten and Schmidt, Fabian",
    title = "{Stellar streams and dark substructure: the diffusion regime}",
    eprint = "2108.13420",
    archivePrefix = "arXiv",
    primaryClass = "astro-ph.GA",
    doi = "10.1093/mnras/stac1022",
    journal = "Mon. Not. Roy. Astron. Soc.",
    volume = "513",
    number = "3",
    pages = "3682--3708",
    year = "2022"
}

@article{Delos:2023dwq,
    author = "Delos, M. Sten",
    title = "{An analytical description of substructure-induced gravitational perturbations in stellar systems}",
    eprint = "2312.13338",
    archivePrefix = "arXiv",
    primaryClass = "astro-ph.GA",
    doi = "10.1093/mnras/stae715",
    journal = "Mon. Not. Roy. Astron. Soc.",
    volume = "529",
    number = "3",
    pages = "2349--2355",
    year = "2024"
}

@article{Xiao:2024qay,
    author = "Xiao, Huangyu and Dai, Liang and McQuinn, Matthew",
    title = "{Detecting dark matter substructures on small scales with fast radio bursts}",
    eprint = "2401.08862",
    archivePrefix = "arXiv",
    primaryClass = "astro-ph.CO",
    reportNumber = "FERMILAB-PUB-24-0004-T",
    doi = "10.1103/PhysRevD.110.023516",
    journal = "Phys. Rev. D",
    volume = "110",
    number = "2",
    pages = "023516",
    year = "2024"
}

@article{Mondino:2020rkn,
    author = "Mondino, Cristina and Taki, Anna-Maria and Van Tilburg, Ken and Weiner, Neal",
    title = "{First Results on Dark Matter Substructure from Astrometric Weak Lensing}",
    eprint = "2002.01938",
    archivePrefix = "arXiv",
    primaryClass = "astro-ph.CO",
    doi = "10.1103/PhysRevLett.125.111101",
    journal = "Phys. Rev. Lett.",
    volume = "125",
    number = "11",
    pages = "111101",
    year = "2020"
}

@article{Cyncynates:2021xzw,
    author = "Cyncynates, David and Giurgica-Tiron, Tudor and Simon, Olivier and Thompson, Jedidiah O.",
    title = "{Resonant nonlinear pairs in the axiverse and their late-time direct and astrophysical signatures}",
    eprint = "2109.09755",
    archivePrefix = "arXiv",
    primaryClass = "hep-ph",
    doi = "10.1103/PhysRevD.105.055005",
    journal = "Phys. Rev. D",
    volume = "105",
    number = "5",
    pages = "055005",
    year = "2022"
}

@article{Zurek:2006sy,
    author = "Zurek, Kathryn M. and Hogan, Craig J. and Quinn, Thomas R.",
    title = "{Astrophysical Effects of Scalar Dark Matter Miniclusters}",
    eprint = "astro-ph/0607341",
    archivePrefix = "arXiv",
    doi = "10.1103/PhysRevD.75.043511",
    journal = "Phys. Rev. D",
    volume = "75",
    pages = "043511",
    year = "2007"
}

@article{Planck:2018vyg,
    author = "Aghanim, N. and others",
    collaboration = "Planck",
    title = "{Planck 2018 results. VI. Cosmological parameters}",
    eprint = "1807.06209",
    archivePrefix = "arXiv",
    primaryClass = "astro-ph.CO",
    doi = "10.1051/0004-6361/201833910",
    journal = "Astron. Astrophys.",
    volume = "641",
    pages = "A6",
    year = "2020"
}

@article{Khlopov:1985jw,
    author = "Khlopov, M. and Malomed, B. A. and Zeldovich, Ia. B.",
    title = "{Gravitational instability of scalar fields and formation of primordial black holes}",
    journal = "Mon. Not. Roy. Astron. Soc.",
    volume = "215",
    pages = "575--589",
    year = "1985"
}

@article{Irsic:2023equ,
    author = "Ir\v{s}i\v{c}, Vid and others",
    title = "{Unveiling dark matter free streaming at the smallest scales with the high redshift Lyman-alpha forest}",
    eprint = "2309.04533",
    archivePrefix = "arXiv",
    primaryClass = "astro-ph.CO",
    doi = "10.1103/PhysRevD.109.043511",
    journal = "Phys. Rev. D",
    volume = "109",
    number = "4",
    pages = "043511",
    year = "2024"
}

@article{Hogan:1988mp,
    author = "Hogan, C.J. and Rees, M.J.",
    title = "{AXION MINICLUSTERS}",
    doi = "10.1016/0370-2693(88)91655-3",
    journal = "Phys. Lett. B",
    volume = "205",
    pages = "228--230",
    year = "1988"
}

@article{Petrossian-Byrne:2025mto,
    author = "Petrossian-Byrne, Rudin and Villadoro, Giovanni",
    title = "{Open String Axiverse}",
    eprint = "2503.16387",
    archivePrefix = "arXiv",
    primaryClass = "hep-ph",
    month = "3",
    year = "2025"
}

@article{Chavanis:2011zm,
    author = "Chavanis, P. H. and Delfini, L.",
    title = "{Mass-radius relation of Newtonian self-gravitating Bose-Einstein condensates with short-range interactions: II. Numerical results}",
    eprint = "1103.2054",
    archivePrefix = "arXiv",
    primaryClass = "astro-ph.CO",
    doi = "10.1103/PhysRevD.84.043532",
    journal = "Phys. Rev. D",
    volume = "84",
    pages = "043532",
    year = "2011"
}

@article{Buckley:2025zgh,
    author = "Buckley, Matthew R. and Du, Peizhi and Fernandez, Nicolas and Weikert, Mitchell J.",
    title = "{General Constraints on Isocurvature from the CMB and Ly-$\alpha$ Forest}",
    eprint = "2502.20434",
    archivePrefix = "arXiv",
    primaryClass = "astro-ph.CO",
    month = "2",
    year = "2025"
}

@article{McQuinn,
    author = "Ir\v{s}i\v{c}, Vid and Xiao, Huangyu and McQuinn, Matthew",
    title = "{Early structure formation constraints on the ultralight axion in the postinflation scenario}",
    eprint = "1911.11150",
    archivePrefix = "arXiv",
    primaryClass = "astro-ph.CO",
    doi = "10.1103/PhysRevD.101.123518",
    journal = "Phys. Rev. D",
    volume = "101",
    number = "12",
    pages = "123518",
    year = "2020"
}

@article{Chung:2023syw,
    author = {Chung, Daniel J. H. and M\"unchmeyer, Moritz and Tadepalli, Sai Chaitanya},
    title = "{Search for isocurvature with large-scale structure: A forecast for Euclid and MegaMapper using EFTofLSS}",
    eprint = "2306.09456",
    archivePrefix = "arXiv",
    primaryClass = "astro-ph.CO",
    doi = "10.1103/PhysRevD.108.103542",
    journal = "Phys. Rev. D",
    volume = "108",
    number = "10",
    pages = "103542",
    year = "2023"
}

@article{Viel:2005qj,
  author         = {Viel, Matteo and Lesgourgues, Julien and Haehnelt, Martin G. and Matarrese, Sabino and Riotto, Antonio},
  title          = {Constraining warm dark matter candidates including sterile neutrinos and light gravitinos with WMAP and the Lyman-alpha forest},
  journal        = {Phys. Rev. D},
  volume         = {71},
  pages          = {063534},
  year           = {2005},
  doi            = {10.1103/PhysRevD.71.063534},
  eprint         = {astro-ph/0501562},
  archivePrefix  = {arXiv},
  primaryClass   = {astro-ph}
}

@article{Ling:2024qfv,
    author = "Ling, Siyang and Amin, Mustafa A.",
    title = "{Free streaming in warm wave dark matter}",
    eprint = "2408.05591",
    archivePrefix = "arXiv",
    primaryClass = "astro-ph.CO",
    doi = "10.1088/1475-7516/2025/02/025",
    journal = "JCAP",
    volume = "02",
    pages = "025",
    year = "2025"
}

@article{Boylan-Kolchin:2022kae,
    author = "Boylan-Kolchin, Michael",
    title = "{Stress testing \ensuremath{\Lambda}CDM with high-redshift galaxy candidates}",
    eprint = "2208.01611",
    archivePrefix = "arXiv",
    primaryClass = "astro-ph.CO",
    doi = "10.1038/s41550-023-01937-7",
    journal = "Nature Astron.",
    volume = "7",
    number = "6",
    pages = "731--735",
    year = "2023"
}

@article{Long:2024imw,
    author = "Long, Andrew J. and Venegas, Moira",
    title = "{Free streaming of warm wave dark matter in modified expansion histories}",
    eprint = "2412.14322",
    archivePrefix = "arXiv",
    primaryClass = "astro-ph.CO",
    month = "12",
    year = "2024"
}

@article{deKruijf:2024voc,
    author = "de Kruijf, Jessie and Vanzan, Eleonora and Boddy, Kimberly K. and Raccanelli, Alvise and Bartolo, Nicola",
    title = "{Searching for blue-tilted power spectra in the dark ages}",
    eprint = "2408.04991",
    archivePrefix = "arXiv",
    primaryClass = "astro-ph.CO",
    reportNumber = "UTWI-26-2024",
    doi = "10.1103/PhysRevD.111.063507",
    journal = "Phys. Rev. D",
    volume = "111",
    number = "6",
    pages = "063507",
    year = "2025"
}

@article{Springel:2020plp,
    author = {Springel, Volker and Pakmor, R\"udiger and Zier, Oliver and Reinecke, Martin},
    title = "{Simulating cosmic structure formation with the gadget-4 code}",
    eprint = "2010.03567",
    archivePrefix = "arXiv",
    primaryClass = "astro-ph.IM",
    doi = "10.1093/mnras/stab1855",
    journal = "Mon. Not. Roy. Astron. Soc.",
    volume = "506",
    number = "2",
    pages = "2871--2949",
    year = "2021"
}

@article{Delos:2018ueo,
    author = "Delos, M. Sten and Erickcek, Adrienne L. and Bailey, Avery P. and Alvarez, Marcelo A.",
    title = "{Density profiles of ultracompact minihalos: Implications for constraining the primordial power spectrum}",
    eprint = "1806.07389",
    archivePrefix = "arXiv",
    primaryClass = "astro-ph.CO",
    doi = "10.1103/PhysRevD.98.063527",
    journal = "Phys. Rev. D",
    volume = "98",
    number = "6",
    pages = "063527",
    year = "2018"
}

@article{Bode:2000gq,
  author         = {Bode, Paul and Ostriker, Jeremiah P. and Turok, Neil},
  title          = {Halo Formation in Warm Dark Matter Models},
  journal        = {Astrophys. J.},
  volume         = {556},
  pages          = {93--107},
  year           = {2001},
  doi            = {10.1086/321541},
  eprint         = {astro-ph/0010389},
  archivePrefix  = {arXiv},
  primaryClass   = {astro-ph}
}

@article{Delos:2024poq,
    author = "Delos, M. Sten and Rantala, Antti and Young, Sam and Schmidt, Fabian",
    title = "{Structure formation with primordial black holes: collisional dynamics, binaries, and gravitational waves}",
    eprint = "2410.01876",
    archivePrefix = "arXiv",
    primaryClass = "astro-ph.CO",
    doi = "10.1088/1475-7516/2024/12/005",
    journal = "JCAP",
    volume = "12",
    pages = "005",
    year = "2024"
}

@article{Inman:2019wvr,
    author = {Inman, Derek and Ali-Ha\"\i{}moud, Yacine},
    title = "{Early structure formation in primordial black hole cosmologies}",
    eprint = "1907.08129",
    archivePrefix = "arXiv",
    primaryClass = "astro-ph.CO",
    doi = "10.1103/PhysRevD.100.083528",
    journal = "Phys. Rev. D",
    volume = "100",
    number = "8",
    pages = "083528",
    year = "2019"
}

@article{Afshordi:2003zb,
    author = "Afshordi, N. and McDonald, P. and Spergel, D. N.",
    title = "{Primordial black holes as dark matter: The Power spectrum and evaporation of early structures}",
    eprint = "astro-ph/0302035",
    archivePrefix = "arXiv",
    doi = "10.1086/378763",
    journal = "Astrophys. J. Lett.",
    volume = "594",
    pages = "L71--L74",
    year = "2003"
}

@article{Delos:2017thv,
    author = "Delos, M. Sten and Erickcek, Adrienne L. and Bailey, Avery P. and Alvarez, Marcelo A.",
    title = "{Are ultracompact minihalos really ultracompact?}",
    eprint = "1712.05421",
    archivePrefix = "arXiv",
    primaryClass = "astro-ph.CO",
    doi = "10.1103/PhysRevD.97.041303",
    journal = "Phys. Rev. D",
    volume = "97",
    number = "4",
    pages = "041303",
    year = "2018"
}

@book{Marsh:2024ury,
    author = "Marsh, David J. E. and Ellis, David and Mehta, Viraf M.",
    title = "{Dark Matter: Evidence, Theory, and Constraints}",
    doi = "10.1515/9780691249711",
    isbn = "978-0-691-24971-1, 978-0-691-24952-0",
    publisher = "Princeton University Press",
    series = "Princeton Series in Astrophysics",
    month = "10",
    year = "2024"
}

@article{Liu:2024pjg,
    author = "Liu, Rayne and Hu, Wayne and Xiao, Huangyu",
    title = "{Warm and fuzzy dark matter: Free streaming of wave dark matter}",
    eprint = "2406.12970",
    archivePrefix = "arXiv",
    primaryClass = "hep-ph",
    reportNumber = "FERMILAB-PUB-24-0296-T",
    doi = "10.1103/PhysRevD.111.023535",
    journal = "Phys. Rev. D",
    volume = "111",
    number = "2",
    pages = "023535",
    year = "2025"
}

@article{Zhou:2024mea,
    author = "Zhou, Shuang-Yong",
    title = "{Non-topological solitons and quasi-solitons}",
    eprint = "2411.16604",
    archivePrefix = "arXiv",
    primaryClass = "hep-th",
    reportNumber = "USTC-ICTS/PCFT-24-51",
    month = "11",
    year = "2024"
}

@article{Delos:2023fpm,
    author = "Delos, M. Sten and Franciolini, Gabriele",
    title = "{Lensing constraints on ultradense dark matter halos}",
    eprint = "2301.13171",
    archivePrefix = "arXiv",
    primaryClass = "astro-ph.CO",
    doi = "10.1103/PhysRevD.107.083505",
    journal = "Phys. Rev. D",
    volume = "107",
    number = "8",
    pages = "083505",
    year = "2023"
}

@article{StenDelos:2022jld,
    author = "Delos, M. Sten and Silk, Joseph",
    title = "{Ultradense dark matter haloes accompany primordial black holes}",
    eprint = "2210.04904",
    archivePrefix = "arXiv",
    primaryClass = "astro-ph.CO",
    doi = "10.1093/mnras/stad356",
    journal = "Mon. Not. Roy. Astron. Soc.",
    volume = "520",
    number = "3",
    pages = "4370--4375",
    year = "2023"
}

@article{Carr:2023tpt,
    author = "Carr, Bernard and Clesse, Sebastien and Garcia-Bellido, Juan and Hawkins, Michael and Kuhnel, Florian",
    title = "{Observational evidence for primordial black holes: A positivist perspective}",
    eprint = "2306.03903",
    archivePrefix = "arXiv",
    primaryClass = "astro-ph.CO",
    doi = "10.1016/j.physrep.2023.11.005",
    journal = "Phys. Rept.",
    volume = "1054",
    pages = "1--68",
    year = "2024"
}

@article{Sabti:2021unj,
    author = "Sabti, Nashwan and Mu{\~n}oz, Julian B. and Blas, Diego",
    title = "{New Roads to the Small-scale Universe: Measurements of the Clustering of Matter with the High-redshift UV Galaxy Luminosity Function}",
    eprint = "2110.13161",
    archivePrefix = "arXiv",
    primaryClass = "astro-ph.CO",
    reportNumber = "KCL-2021-75",
    doi = "10.3847/2041-8213/ac5e9c",
    journal = "Astrophys. J. Lett.",
    volume = "928",
    number = "2",
    pages = "L20",
    year = "2022"
}

@article{Gilman:2021gkj,
    author = "Gilman, Daniel and Benson, Andrew and Bovy, Jo and Birrer, Simon and Treu, Tommaso and Nierenberg, Anna",
    title = "{The primordial matter power spectrum on sub-galactic scales}",
    eprint = "2112.03293",
    archivePrefix = "arXiv",
    primaryClass = "astro-ph.CO",
    doi = "10.1093/mnras/stac670",
    journal = "Mon. Not. Roy. Astron. Soc.",
    volume = "512",
    number = "3",
    pages = "3163--3188",
    year = "2022"
}

@article{Esteban:2023xpk,
    author = "Esteban, Ivan and Peter, Annika H. G. and Kim, Stacy Y.",
    title = "{Milky~Way satellite velocities reveal the dark matter power spectrum at small scales}",
    eprint = "2306.04674",
    archivePrefix = "arXiv",
    primaryClass = "astro-ph.CO",
    doi = "10.1103/PhysRevD.110.123013",
    journal = "Phys. Rev. D",
    volume = "110",
    number = "12",
    pages = "123013",
    year = "2024"
}

@article{Boddy:2025oxn,
    author = "Boddy, Kimberly K. and Dror, Jeff A. and Lam, Austin",
    title = "{Ultralight Dark Matter Statistics for Pulsar Timing Detection}",
    eprint = "2502.15874",
    archivePrefix = "arXiv",
    primaryClass = "hep-ph",
    month = "2",
    year = "2025"
}

@article{Liu:2025lts,
    author = "Liu, Rayne and Hu, Wayne and Xiao, Huangyu",
    title = "{Interference with Gravitational Instability: Hot and Fuzzy Dark Matter}",
    eprint = "2504.01937",
    archivePrefix = "arXiv",
    primaryClass = "astro-ph.CO",
    reportNumber = "FERMILAB-PUB-25-0204-T",
    month = "4",
    year = "2025"
}

@article{Amin:2025dtd,
    author = "Amin, Mustafa A. and Delos, M. Sten and Mirbabayi, Mehrdad",
    title = "{Structure Formation with Warm White Noise: Effects of Finite Number Density and Velocity Dispersion in Particle and Wave Dark Matter}",
    eprint = "2503.20881",
    archivePrefix = "arXiv",
    primaryClass = "astro-ph.CO",
    month = "3",
    year = "2025"
}

@article{Amin:2025sla,
    author = "Amin, Mustafa A. and May, Simon and Mirbabayi, Mehrdad",
    title = "{Early Growth of Structure in Warm Wave Dark Matter}",
    eprint = "2506.12131",
    archivePrefix = "arXiv",
    primaryClass = "astro-ph.CO",
    month = "6",
    year = "2025"
}

@article{Harigaya:2025pox,
    author = "Harigaya, Keisuke and Hu, Wayne and Liu, Rayne and Xiao, Huangyu",
    title = "{Universal lower bound on the axion decay constant from free streaming effects}",
    eprint = "2507.01956",
    archivePrefix = "arXiv",
    primaryClass = "astro-ph.CO",
    reportNumber = "FERMILAB-PUB-25-0430-T",
    month = "7",
    year = "2025"
}

@article{Kaup:1968boson,
  author       = {Kaup, Donald J.},
  title        = {Klein--Gordon Geon},
  journal      = {Physical Review},
  volume       = {172},
  number       = {5},
  pages        = {1331--1342},
  year         = {1968},
  doi          = {10.1103/PhysRev.172.1331}
}

@article{Zeldovich:1967pbh,
  author       = {Zeldovich, Ya. B. and Novikov, I. D.},
  title        = {The Hypothesis of Cores Retarded during Expansion and the Hot Cosmological Model},
  journal      = {Soviet Astronomy},
  volume       = {10},
  pages        = {602--603},
  year         = {1967},
  note         = {Originally published in Astronomicheskii Zhurnal, Vol. 43, p. 758 (1966)}
}

@article{Ling:2025ehm,
    author = "Ling, Siyang",
    title = "{Generating Moving Field Initial Conditions with Spatially Varying Boost}",
    eprint = "2506.23020",
    archivePrefix = "arXiv",
    primaryClass = "physics.comp-ph",
    month = "6",
    year = "2025"
}

@article{Alexander:2024nvi,
    author = "Alexander, Stephon and Manton, Tucker and McDonough, Evan",
    title = "{Field theory axiverse}",
    eprint = "2404.11642",
    archivePrefix = "arXiv",
    primaryClass = "hep-ph",
    doi = "10.1103/PhysRevD.109.116019",
    journal = "Phys. Rev. D",
    volume = "109",
    number = "11",
    pages = "116019",
    year = "2024"
}

@article{Viel:2013apy,
  author         = {Viel, Matteo and Becker, George D. and Bolton, James S. and Haehnelt, Martin G.},
  title          = {Warm dark matter as a solution to the small scale crisis: New constraints from high redshift Lyman-alpha forest data},
  journal        = {Phys. Rev. D},
  volume         = {88},
  pages          = {043502},
  year           = {2013},
  doi            = {10.1103/PhysRevD.88.043502},
  eprint         = {1306.2314},
  archivePrefix  = {arXiv},
  primaryClass   = {astro-ph.CO}
}

@article{Graham:2025fdt,
    author = "Graham, Peter W. and Green, Daniel and Meyers, Joel",
    title = "{Dark Forces Gathering}",
    eprint = "2508.20999",
    archivePrefix = "arXiv",
    primaryClass = "astro-ph.CO",
    month = "8",
    year = "2025"
}

@article{Celik:2025wkt,
    author = "{\c{C}}elik, {\c{S}}afak and Schmidt, Fabian",
    title = "{Mixed Dark Matter and Galaxy Clustering: The Importance of Relative Perturbations}",
    eprint = "2508.21481",
    archivePrefix = "arXiv",
    primaryClass = "astro-ph.CO",
    month = "8",
    year = "2025"
}

@article{Bottaro:2023wkd,
    author = "Bottaro, Salvatore and Castorina, Emanuele and Costa, Marco and Redigolo, Diego and Salvioni, Ennio",
    title = "{Unveiling Dark Forces with Measurements of the Large Scale Structure of the Universe}",
    eprint = "2309.11496",
    archivePrefix = "arXiv",
    primaryClass = "astro-ph.CO",
    doi = "10.1103/PhysRevLett.132.201002",
    journal = "Phys. Rev. Lett.",
    volume = "132",
    number = "20",
    pages = "201002",
    year = "2024"
}

@article{Bottaro:2024pcb,
    author = "Bottaro, Salvatore and Castorina, Emanuele and Costa, Marco and Redigolo, Diego and Salvioni, Ennio",
    title = "{From 100~kpc to 10~Gpc: Dark matter self-interactions before and after DESI observations}",
    eprint = "2407.18252",
    archivePrefix = "arXiv",
    primaryClass = "astro-ph.CO",
    doi = "10.1103/gc78-96l5",
    journal = "Phys. Rev. D",
    volume = "112",
    number = "2",
    pages = "023525",
    year = "2025"
}

@article{Lague:2023wes,
    author = {Lagu{\"e}, Alex and Schwabe, Bodo and Hlo{\v{z}}ek, Ren{\'e}e and Marsh, David J. E. and Rogers, Keir K.},
    title = "{Cosmological simulations of mixed ultralight dark matter}",
    eprint = "2310.20000",
    archivePrefix = "arXiv",
    primaryClass = "astro-ph.CO",
    reportNumber = "KCL-PH-TH/2023-57",
    doi = "10.1103/PhysRevD.109.043507",
    journal = "Phys. Rev. D",
    volume = "109",
    number = "4",
    pages = "043507",
    year = "2024"
}

@article{Tan:2024cek,
    author = "Tan, Chin Yi and Dekker, Ariane and Drlica-Wagner, Alex",
    title = "{Mixed warm dark matter constraints using Milky~Way satellite galaxy counts}",
    eprint = "2409.18917",
    archivePrefix = "arXiv",
    primaryClass = "astro-ph.CO",
    reportNumber = "FERMILAB-PUB-24-0586-PPD",
    doi = "10.1103/PhysRevD.111.063079",
    journal = "Phys. Rev. D",
    volume = "111",
    number = "6",
    pages = "063079",
    year = "2025"
}

@article{Garcia-Gallego:2025kiw,
    author = "Garcia-Gallego, Olga and Ir{\v{s}}i{\v{c}}, Vid and Haehnelt, Martin G. and Viel, Matteo and Bolton, James S.",
    title = "{Constraining mixed dark matter models with high-redshift Lyman-alpha forest data}",
    eprint = "2504.06367",
    archivePrefix = "arXiv",
    primaryClass = "astro-ph.CO",
    doi = "10.1103/4k29-h99l",
    journal = "Phys. Rev. D",
    volume = "112",
    number = "4",
    pages = "043502",
    year = "2025"
}

@article{Garny:2022kbk,
    author = "Garny, Mathias and Laxhuber, Dominik and Scoccimarro, Roman",
    title = "{Perturbation theory with dispersion and higher cumulants: Nonlinear regime}",
    eprint = "2210.08089",
    archivePrefix = "arXiv",
    primaryClass = "astro-ph.CO",
    reportNumber = "TUM-HEP 1424/22",
    doi = "10.1103/PhysRevD.107.063540",
    journal = "Phys. Rev. D",
    volume = "107",
    number = "6",
    pages = "063540",
    year = "2023"
}

@article{Garny:2022tlk,
    author = "Garny, Mathias and Laxhuber, Dominik and Scoccimarro, Roman",
    title = "{Perturbation theory with dispersion and higher cumulants: Framework and linear theory}",
    eprint = "2210.08088",
    archivePrefix = "arXiv",
    primaryClass = "astro-ph.CO",
    reportNumber = "TUM-HEP 1423/22",
    doi = "10.1103/PhysRevD.107.063539",
    journal = "Phys. Rev. D",
    volume = "107",
    number = "6",
    pages = "063539",
    year = "2023"
}

@article{Carr:2018rid,
    author = "Carr, Bernard and Silk, Joseph",
    title = "{Primordial Black Holes as Generators of Cosmic Structures}",
    eprint = "1801.00672",
    archivePrefix = "arXiv",
    primaryClass = "astro-ph.CO",
    doi = "10.1093/mnras/sty1204",
    journal = "Mon. Not. Roy. Astron. Soc.",
    volume = "478",
    number = "3",
    pages = "3756--3775",
    year = "2018"
}

@article{Zurek:2008qg,
    author = "Zurek, Kathryn M.",
    title = "{Multi-Component Dark Matter}",
    eprint = "0811.4429",
    archivePrefix = "arXiv",
    primaryClass = "hep-ph",
    reportNumber = "FERMILAB-PUB-08-542-A",
    doi = "10.1103/PhysRevD.79.115002",
    journal = "Phys. Rev. D",
    volume = "79",
    pages = "115002",
    year = "2009"
}

@article{Primack1997,
  author       = {Primack, Joel R.},
  title        = {Status of Cold + Hot Dark Matter Cosmogony},
  journal      = {Nuclear Physics B - Proceedings Supplements},
  volume       = {51},
  pages        = {30--41},
  year         = {1996},
  doi          = {10.1016/S0920-5632(96)00646-2},
  eprint       = {astro-ph/9707285},
  archivePrefix= {arXiv},
  primaryClass = {astro-ph},
  note         = {Historical review of CHDM models}
}

@article{Vanzan:2023gui,
    author = "Vanzan, Eleonora and Raccanelli, Alvise and Bartolo, Nicola",
    title = "{Dark ages, a window on the dark sector. Hunting for ultra-light axions}",
    eprint = "2306.09252",
    archivePrefix = "arXiv",
    primaryClass = "astro-ph.CO",
    doi = "10.1088/1475-7516/2024/03/001",
    journal = "JCAP",
    volume = "03",
    pages = "001",
    year = "2024"
}

@article{Verdiani:2025jcf,
    author = "Verdiani, Francesco and Castorina, Emanuele and Salvioni, Ennio and Sefusatti, Emiliano",
    title = "{The Effective Field Theory of Large Scale Structure for Mixed Dark Matter Scenarios}",
    eprint = "2507.08792",
    archivePrefix = "arXiv",
    primaryClass = "astro-ph.CO",
    month = "7",
    year = "2025"
}

@article{Cicoli:2023opf,
    author = "Cicoli, Michele and Conlon, Joseph P. and Maharana, Anshuman and Parameswaran, Susha and Quevedo, Fernando and Zavala, Ivonne",
    title = "{String cosmology: From the early universe to today}",
    eprint = "2303.04819",
    archivePrefix = "arXiv",
    primaryClass = "hep-th",
    doi = "10.1016/j.physrep.2024.01.002",
    journal = "Phys. Rept.",
    volume = "1059",
    pages = "1--155",
    year = "2024"
}

@article{Chabanier:2019eai,
    author = "Chabanier, Sol{\`e}ne and Millea, Marius and Palanque-Delabrouille, Nathalie",
    title = "{Matter power spectrum: from Ly$\alpha$ forest to CMB scales}",
    eprint = "1905.08103",
    archivePrefix = "arXiv",
    primaryClass = "astro-ph.CO",
    reportNumber = "Volume: 489, pages = 2247-2253",
    doi = "10.1093/mnras/stz2310",
    journal = "Mon. Not. Roy. Astron. Soc.",
    volume = "489",
    number = "2",
    pages = "2247--2253",
    year = "2019"
}

@article{An:2025gju,
    author = "An, Rui and Nadler, Ethan O. and Benson, Andrew and Gluscevic, Vera",
    title = "{COZMIC. II. Cosmological Zoom-in Simulations with Fractional non-CDM Initial Conditions}",
    eprint = "2411.03431",
    archivePrefix = "arXiv",
    primaryClass = "astro-ph.CO",
    doi = "10.3847/1538-4357/adce83",
    journal = "Astrophys. J.",
    volume = "986",
    pages = "128",
    year = "2025"
}

@article{Fernandez:2023ddy,
    author = "Fernandez, Nicolas and Foster, Joshua W. and Lillard, Benjamin and Shelton, Jessie",
    title = "{Stochastic Gravitational Waves from Early Structure Formation}",
    eprint = "2312.12499",
    archivePrefix = "arXiv",
    primaryClass = "astro-ph.CO",
    reportNumber = "MIT-CTP/5660",
    doi = "10.1103/PhysRevLett.133.111002",
    journal = "Phys. Rev. Lett.",
    volume = "133",
    number = "11",
    pages = "111002",
    year = "2024"
}

@article{Trost:2024ciu,
    author = "Trost, Andrea and Bolton, James S. and Caputo, Andrea and Liu, Hongwan and Cristiani, Stefano and Viel, Matteo",
    title = "{Constraints on dark photon dark matter from Lyman-{\ensuremath{\alpha}} forest simulations and an ultrahigh signal-to-noise quasar spectrum}",
    eprint = "2410.02858",
    archivePrefix = "arXiv",
    primaryClass = "astro-ph.CO",
    reportNumber = "FERMILAB-PUB-24-0739-V, CERN-TH-2024-166",
    doi = "10.1103/PhysRevD.111.083034",
    journal = "Phys. Rev. D",
    volume = "111",
    number = "8",
    pages = "083034",
    year = "2025"
}

@article{Schwabe:2020eac,
    author = "Schwabe, Bodo and Gosenca, Mateja and Behrens, Christoph and Niemeyer, Jens C. and Easther, Richard",
    title = "{Simulating mixed fuzzy and cold dark matter}",
    eprint = "2007.08256",
    archivePrefix = "arXiv",
    primaryClass = "astro-ph.CO",
    doi = "10.1103/PhysRevD.102.083518",
    journal = "Phys. Rev. D",
    volume = "102",
    number = "8",
    pages = "083518",
    year = "2020"
}

@article{DESI:2024mwx,
    author = "Adame, A. G. and others",
    collaboration = "DESI",
    title = "{DESI 2024 VI: cosmological constraints from the measurements of baryon acoustic oscillations}",
    eprint = "2404.03002",
    archivePrefix = "arXiv",
    primaryClass = "astro-ph.CO",
    reportNumber = "FERMILAB-PUB-24-0154-PPD",
    doi = "10.1088/1475-7516/2025/02/021",
    journal = "JCAP",
    volume = "02",
    pages = "021",
    year = "2025"
}

@article{DES:2021wwk,
    author = "Abbott, T. M. C. and others",
    collaboration = "DES",
    title = "{Dark Energy Survey Year 3 results: Cosmological constraints from galaxy clustering and weak lensing}",
    eprint = "2105.13549",
    archivePrefix = "arXiv",
    primaryClass = "astro-ph.CO",
    reportNumber = "FERMILAB-PUB-21-221-AE, DES-2020-0617",
    doi = "10.1103/PhysRevD.105.023520",
    journal = "Phys. Rev. D",
    volume = "105",
    number = "2",
    pages = "023520",
    year = "2022"
}

@article{Lesgourgues:2012uu,
    author = "Lesgourgues, Julien and Pastor, Sergio",
    title = "{Neutrino mass from Cosmology}",
    eprint = "1212.6154",
    archivePrefix = "arXiv",
    primaryClass = "hep-ph",
    doi = "10.1155/2012/608515",
    journal = "Adv. High Energy Phys.",
    volume = "2012",
    pages = "608515",
    year = "2012"
}

@article{Chudaykin:2020aoj,
    author = "Chudaykin, Anton and Ivanov, Mikhail M. and Philcox, Oliver H. E. and Simonovi{\'c}, Marko",
    title = "{Nonlinear perturbation theory extension of the Boltzmann code CLASS}",
    eprint = "2004.10607",
    archivePrefix = "arXiv",
    primaryClass = "astro-ph.CO",
    reportNumber = "INR-TH-2020-016, CERN-TH-2020-062",
    doi = "10.1103/PhysRevD.102.063533",
    journal = "Phys. Rev. D",
    volume = "102",
    number = "6",
    pages = "063533",
    year = "2020"
}

@article{Tadepalli:2025gzf,
    author = "Tadepalli, Sai Chaitanya and Takahashi, Tomo",
    title = "{Warm Dark Matter meets Cold Dark Matter Isocurvature}",
    eprint = "2508.03805",
    archivePrefix = "arXiv",
    primaryClass = "astro-ph.CO",
    month = "8",
    year = "2025"
}

@article{Ivanov:2025pbu,
    author = "Ivanov, Mikhail M. and Trifinopoulos, Sokratis",
    title = "{Effective Field Theory Constraints on Primordial Black Holes from the High-Redshift Lyman-$\alpha$ Forest}",
    eprint = "2508.04767",
    archivePrefix = "arXiv",
    primaryClass = "astro-ph.CO",
    reportNumber = "MIT-CTP/5895, CERN-TH-2025-155",
    month = "8",
    year = "2025"
}

@article{Gerlach:2025vco,
    author = "Gerlach, Christopher and Gouttenoire, Yann and Iovino, Antonio J. and Leister, Nicholas",
    title = "{Closing the Mass Window for Stupendously Large Black Holes}",
    eprint = "2508.08238",
    archivePrefix = "arXiv",
    primaryClass = "astro-ph.CO",
    reportNumber = "MITP-25-053",
    month = "8",
    year = "2025"
}

@article{Amin:2025nxm,
    author = "Amin, Mustafa A. and Delos, M. Sten",
    title = "{Growth of Structure in Multi-species Wave Dark Matter}",
    eprint = "2510.17977",
    archivePrefix = "arXiv",
    primaryClass = "astro-ph.CO",
    month = "10",
    year = "2025"
}

@article{Jenkins:2020ctp,
    author = "Jenkins, Alexander C. and Sakellariadou, Mairi",
    title = "{Primordial black holes from cusp collapse on cosmic strings}",
    eprint = "2006.16249",
    archivePrefix = "arXiv",
    primaryClass = "astro-ph.CO",
    reportNumber = "KCL-PH-TH/2020-30",
    month = "6",
    year = "2020"
}

@article{Yuwen:2024gcf,
    author = "Yuwen, Zi-Yan and Joana, Cristian and Wang, Shao-Jiang and Cai, Rong-Gen",
    title = "{Bubbles kick off primordial black holes to form more binaries}",
    eprint = "2406.05838",
    archivePrefix = "arXiv",
    primaryClass = "gr-qc",
    doi = "10.1103/PhysRevResearch.7.023180",
    journal = "Phys. Rev. Res.",
    volume = "7",
    number = "2",
    pages = "023180",
    year = "2025"
}

@article{Wang:2025mea,
    author = "Wang, Xinpeng and Lu, Yifan and Picker, Zachary S. C. and Kusenko, Alexander and Sasaki, Misao",
    title = "{When Tiny Halos Stir Spacetime: Gravitational Waves from Fifth-Force Mergers}",
    eprint = "2510.12984",
    archivePrefix = "arXiv",
    primaryClass = "astro-ph.CO",
    month = "10",
    year = "2025"
}
\appendix

\section{From Liouville Equation to Transfer Functions}
\label{App:Derivation}

In this appendix, we derive the main results presented in Section~\ref{sec:Main}. Similarly to \cite{Amin:2025dtd}, we begin by analyzing perturbations in a homogeneous static universe for simplicity.\footnote{Technically, the analysis of perturbations to a static background involves a ``Jeans swindle'' \cite{Binney:1987}, since we neglect that the background itself would collapse under the gravity of the mass distribution. This problem vanishes when we translate the results to the expanding-universe case.} In Appendix~\ref{App:Expanding}, we finally convert the results to the case of an expanding universe.

\subsection{The 2-particle probability distribution}

The distribution function $F=F(t,\bx_1,\bx_2,...,\bv_1,\bv_2,...)$ is the joint probability distribution for the phase-space positions of all particles in the system. It evolves according to the Liouville equation,
\beq
\partial_t F
=
-\sum_i\bv_i\cdot \nabla_{\bx_i} F
-\sum_i\sum_{j\neq i}\ba_{ij}\cdot \nabla_{\bv_i} F,
\qquad
\ba_{ij}=-Gm_j \frac{(\bx_i-\bx_j)}{|\bx_i-\bx_j|^3},
\eeq
where $\ba_{ij}$ is the contribution to the acceleration of particle $i$ due to particle $j$.
The phase-space position of a particular particle $s$ has the probability distribution\footnote{In \cite{Amin:2025dtd}, $f^{(s)}$ (with the braces in the exponent) referred to an $s$-particle distribution function. Here $f^s$ is the one particle distribution function for particle labeled by index $s$. Moreover, here $\int \dl\bx_s \dl\bv_s f^s(t,\bx_s,\bv_s)=1$, whereas in \cite{Amin:2025dtd}, we chose $V^{-1}\int \dl\bp \dl\bx f^{(1)}(t,\bx,\bp)=1$. Furthermore, where we define the one particle distribution function for a species $S$, $f^S$, we will normalize it as $\int \dl \bv f^S=1$.}
\beq
f^s(t,\bx_s,\bv_s)
&=
\int\left(\prod_{s'\neq s}\di\bx_{s'}\di\bv_{s'}\right)F.
\eeq
By integrating the Liouville equation, one can show that $f^s$ evolves according to
\beq\label{eq:f1eq}
\partial_t f^s
&=
-\bv_s\cdot \nabla_{\bx_s} f^s
-\sum_{s'\neq s}\int\di\bx_{s'}\di\bv_{s'}\ba_{ss'}\cdot \nabla_{\bv_s} f^{ss'},
\eeq
where $f^{ss'}$ is the joint distribution
\beq
f^{ss'}(t,\bx_s,\bx_{s'},\bv_s,\bv_{s'})
&=
\int\left(\prod_{s''\neq s,s'}\di\bx_{s''}\di\bv_{s''}\right)F.
\eeq
A corresponding integration of the Liouville equation reveals that
\beq\label{eq:f2eq}
\partial_t f^{ss'}
&=
-\bv_s\cdot \nabla_{\bx_s} f^{ss'}
-\bv_{s'}\cdot \nabla_{\bx_{s'}} f^{ss'}
-\ba_{ss'}\cdot \nabla_{\bv_s} f^{ss'}
-\ba_{s's}\cdot \nabla_{\bv_{s'}} f^{ss'}
\\&\hphantom{=}-\sum_{s''\neq s,s'}\int\di\bx_{s''}\di\bv_{s''}\ba_{ss''}\cdot \nabla_{\bv_s} f^{ss's''}
-\sum_{s''\neq s,s'}\int\di\bx_{s''}\di\bv_{s''}\ba_{s's''}\cdot \nabla_{\bv_{s'}} f^{ss's''},
\eeq
where $f^{ss's''}$ is similarly the joint distribution of the phase-space positions of the three particles $s$, $s'$, and $s''$.

Now write the Mayer cluster expansion \cite{mayer1948statistical}
\beq
f^{ss'} &= f^s f^{s'} + g^{ss'},\\
f^{ss's''} &= f^s f^{s'} f^{s''} + f^s g^{s's''} + f^{s'} g^{ss''} + f^{s''} g^{ss'} + h^{ss's''}.
\eeq
By substituting the expansion of $f^{ss'}$ into equation~\eqref{eq:f1eq}, we find that $f^s$ evolves according to
\beq\label{eq:f1geq}
\partial_t f^s
&=
-\bv_s\cdot\nabla_{\bx_s} f^s
-\left(\sum_{s'\neq s}\int\di\bx_{s'}\di\bv_{s'}\ba_{ss'}f^{s'}\right)\cdot\nabla_{\bv_s}f^s
-\sum_{s'\neq s}\int\di\bx_{s'}\di\bv_{s'}\ba_{ss'} \cdot\nabla_{\bv_s} g^{ss'}.
\eeq
Meanwhile, by substituting both expansions into equation~\eqref{eq:f2eq} and using equation~\eqref{eq:f1geq} to simplify, we find that $g^{ss'}$ evolves as
\beq
\partial_t g^{ss'}
&=
f^s\nabla_{\bv_{s'}}f^{s'}\cdot\int\di\bx\di\bv\ba_{s's}(\bx_{s'},\bx)f^{s}(t,\bx,\bv)
+f^{s'}\nabla_{\bv_s}f^s\cdot\int\di\bx\di\bv\ba_{ss'}(\bx_s,\bx)f^{s'}(t,\bx,\bv)
\\&\hphantom{=}
+f^s\int\di\bx\di\bv\ba_{s's}(\bx_{s'},\bx) \cdot\nabla_{\bv_{s'}} g^{s's}(t,\bx_{s'},\bx,\bv_{s'},\bv)
\\&\hphantom{=}
+f^{s'}\int\di\bx\di\bv\ba_{ss'}(\bx_s,\bx) \cdot\nabla_{\bv_s} g^{ss'}(t,\bx_s,\bx,\bv_s,\bv)
\\&\hphantom{=}
-f^s\ba_{s's}\cdot \nabla_{\bv_{s'}} f^{s'}
-f^{s'}\ba_{ss'}\cdot \nabla_{\bv_s} f^s
\\&\hphantom{=}
-\bv_s\cdot \nabla_{\bx_s} g^{ss'}
-\bv_{s'}\cdot \nabla_{\bx_{s'}} g^{ss'}
-\ba_{ss'}\cdot \nabla_{\bv_s} g^{ss'}
-\ba_{s's}\cdot \nabla_{\bv_{s'}} g^{ss'}
\\&\hphantom{=}
-\nabla_{\bv_s} f^s \cdot\sum_{s''\neq s,s'}\int\di\bx_{s''}\di\bv_{s''}\ba_{ss''}g^{s's''}
-\nabla_{\bv_{s'}} f^{s'} \cdot\sum_{s''\neq s,s'}\int\di\bx_{s''}\di\bv_{s''}\ba_{s's''}g^{ss''}
\\&\hphantom{=}
-\left(\sum_{s''\neq s,s'}\int\di\bx_{s''}\di\bv_{s''}\ba_{ss''}f^{s''} \right)\cdot \nabla_{\bv_s} g^{ss'}
-\left(\sum_{s''\neq s,s'}\int\di\bx_{s''}\di\bv_{s''}\ba_{s's''} f^{s''} \right)\cdot \nabla_{\bv_{s'}}g^{ss'}
\\&\hphantom{=}
-\sum_{s''\neq s,s'}\int\di\bx_{s''}\di\bv_{s''}\ba_{ss''}\cdot \nabla_{\bv_s} h^{ss's''}
-\sum_{s''\neq s,s'}\int\di\bx_{s''}\di\bv_{s''}\ba_{s's''}\cdot \nabla_{\bv_{s'}} h^{ss's''}.
\eeq
In all equations above, unless otherwise specified, $f^i=f^i(t,\bx_i,\bv_i)$, $g^{ij}=g^{ij}(t,\bx_i,\bx_{j},\bv_i,\bv_{j})$, and $\ba_{ij}=\ba_{ij}(\bx_i,\bx_j)$. However, henceforth we write out the function arguments explicitly.
Let us assume spatial homogeneity, so that terms of the form $\int\di\bx\,\ba f$ vanish. Let us also approximate $g^{ss'}\ll f^s f^{s'}$ and $h^{ss's''}\ll f^s g^{ss'}$ (and permutations thereof). We will also ignore the time evolution of $f^s$, that is $f^s(t,\bv)=f^s(\bv)$ which is a reasonable assumption until nonlinear clustering begins, see discussion in \cite{Amin:2025dtd} and also in \cite{Garny:2022tlk,Garny:2022kbk}. Additionally neglecting spatially homogeneous terms, we arrive at
\beq
\partial_t g^{ss'}(t,\bx,\bx',\bv,\bv')
&=
-f^s(\bv)\ba_{s's}(\bx',\bx)\cdot \nabla_{\bv'} f^{s'}(\bv')
-f^{s'}(\bv')\ba_{ss'}(\bx,\bx')\cdot \nabla_{\bv} f^s(\bv)
\\&\hphantom{=}
-\bv\cdot \nabla_{\bx} g^{ss'}(t,\bx,\bx',\bv,\bv')
-\bv'\cdot \nabla_{\bx'} g^{ss'}(t,\bx,\bx',\bv,\bv')
\\&\hphantom{=}
-\nabla_{\bv} f^s(\bv) \cdot\sum_{s''\neq s,s'}\int\di\bx''\di\bv''\ba_{ss''}(\bx,\bx'')g^{s's''}(t,\bx',\bx'',\bv',\bv'')
\\&\hphantom{=}
-\nabla_{\bv'} f^{s'}(\bv') \cdot\sum_{s''\neq s,s'}\int\di\bx''\di\bv''\ba_{s's''}(\bx',\bx'')g^{ss''}(t,\bx,\bx'',\bv,\bv'').
\eeq
In Fourier space, this equation becomes
\beq
\partial_t g_{\bk\bk'}^{ss'}(t,\bv,\bv')
&=
-4\pi G \ddelta(\bk+\bk')\left[m_sf^s(\bv) \frac{i\bk'}{k'^2}\cdot \nabla_{\bv'} f^{s'}(\bv')
+
m_{s'}f^{s'}(\bv') \frac{i\bk}{k^2}\cdot \nabla_{\bv} f^s(\bv)\right]
\\&\hphantom{=}
-i\bk\cdot\bv g_{\bk\bk'}^{ss'}(t,\bv,\bv')
-i\bk'\cdot\bv' g_{\bk\bk'}^{ss'}(t,\bv,\bv')
\\&\hphantom{=}
-4\pi G\frac{i\bk}{k^2}\cdot\nabla_{\bv} f^s(\bv)\sum_{s''\neq s,s'}\int\di\bv'' m_{s''} g_{\bk'\bk}^{s's''}(t,\bv',\bv'')
\\&\hphantom{=}
-4\pi G\frac{i\bk'}{k'^2}\cdot\nabla_{\bv'} f^{s'}(\bv') \sum_{s''\neq s,s'}\int\di\bv'' m_{s''} g_{\bk\bk'}^{ss''}(t,\bv,\bv'').
\eeq


To solve the equation for $g^{ss'}$, let us define the functions
\beq\label{eq:gamma_abc}
\gamma_\bk^{(\mathrm{a})s}(\bv)=f^s(\bv),
\qquad
\gamma_\bk^{(\mathrm{b})s}(\bv)=-\frac{i\bk}{k^2}\cdot\nabla_{\bv} f^s(\bv),
\qquad
\gamma_\bk^{(\mathrm{c})s}(\bv)=\frac{\overline{m}\,m_s}{\overline{m^2}} f^s(\bv).
\eeq
For later convenience, we include in the $\gamma_\bk^{(\mathrm{c})s}$ definition the ratio $\overline{m}/\overline{m^2}=(\sum_{s'}m_{s'})/(\sum_{s'}m_{s'}^2)$ between the mean particle mass and the mean squared particle mass.
Note that $f^s$ has dimensions of $(\text{velocity})^{-3}(\text{length})^{-3}$, so $\gamma^{(\mathrm{a})s}$ and $\gamma^{(\mathrm{c})s}$ have those same dimensions, and $\gamma^{(\mathrm{b})s}$ has dimensions of $(\text{velocity})^{-3}(\text{length})^{-3}(\text{time})$.
In these terms, the two-particle correlation function $g^{ss'}$ evolves according to
\beq\label{eq:gss_evo}
&
\partial_t g_{\bk\bk'}^{ss'}(t,\bv,\bv')
+i\bk\cdot\bv g_{\bk\bk'}^{ss'}(t,\bv,\bv')
+i\bk'\cdot\bv' g_{\bk\bk'}^{ss'}(t,\bv,\bv')
\\&
-4\pi G\gamma_\bk^{(\mathrm{b})s}(\bv)\sum_{s''\neq s,s'}\int\di\bv'' m_{s''} g_{\bk'\bk}^{s's''}(t,\bv',\bv'')
-4\pi G\gamma_{\bk'}^{(\mathrm{b})s'}(\bv')\sum_{s''\neq s,s'}\int\di\bv'' m_{s''} g_{\bk\bk'}^{ss''}(t,\bv,\bv'')
\\&
=
4\pi G\frac{\overline{m^2}}{\overline{m}}\ddelta(\bk+\bk')
\left[\gamma_\bk^{(\mathrm{c})s}(\bv)\gamma_{\bk'}^{(\mathrm{b})s'}(\bv')
+
\gamma_{\bk'}^{(\mathrm{c})s'}(\bv')\gamma_\bk^{(\mathrm{b})s}(\bv)\right].
\eeq

\paragraph{Source-free solution:}
The homogeneous version of equation~\eqref{eq:gss_evo} (with the right-hand side taken to be 0) is solved by expressions of the form
\beq
g^{ss'}_{\bk\bk'}(t,\bv,\bv')
&= \gamma_\bk^{(i)s}(t,\bv)\gamma_{\bk'}^{(j)s'}(t,\bv')\ddelta(\bk+\bk')
\eeq
and sums thereof, where the $\gamma_\bk^{(i)s}(t,\bv)$ are functions that satisfy the equation
\beq\label{eq:gamma_i_eq}
\partial_t \gamma_\bk^{(i)s}(t,\bv)
+i\bv\cdot\bk\gamma_\bk^{(i)s}(t,\bv)
-4\pi G \gamma_\bk^{(\mathrm{b})s}(\bv)\int \di\bv'\sum_{s'}m_{s'}\gamma_\bk^{(i)s'}(t,\bv')
=0.
\eeq
For an initial condition $\gamma_\bk^{(i)s}(t_0,\bv)=\gamma_\bk^{(i)s}(\bv)$ at time $t=t_0$, the solution to this equation is
\beq\label{eq:gamma_i_sol}
\gamma^{(i)s}_{\bk}(t,t_0,\bv)
=e^{-i\bk\cdot\bv (t-t_0)}\gamma^{(i)s}_{\bk}(\bv)
&+4\pi G\int_{t_0}^t dt'e^{-i\bk\cdot\bv (t-t')} \gamma_\bk^{(\mathrm{b})s}(\bv)\int \di\bv'\sum_{s'}m_{s'}\gamma^{(i)s'}_{\bk}(t',t_0,\bv').
\eeq
Here ``$(i)$'' is an arbitrary label for the function. However we will be especially interested in the solutions $(i)=(\mathrm{a})$, $(\mathrm{b})$, $(\mathrm{c})$ with initial conditions corresponding to equations~\eqref{eq:gamma_abc}.

In particular, we will be interested in adiabatic initial perturbations, which correspond to the homogeneous solutions
\beq\label{eq:gss_ad}
g^{ss'}_{\bk\bk'}(t,\bv,\bv')
&=
P^{(\mathrm{ad})}_\delta(t_0,k)
\gamma_\bk^{(\mathrm{ad})s}(t,t_0,\bv)\gamma_{\bk'}^{(\mathrm{ad})s'}(t,t_0,\bv')\ddelta(\bk+\bk'),
\eeq
where $P^{(\mathrm{ad})}_\delta(t_0,k)$ is the power spectrum of adiabatic density perturbations at the initial time $t_0$ and
\beq\label{eq:gamma_ad}
\gamma_\bk^{(\mathrm{ad})s}(t,t_0,\bv)
&=
\gamma_\bk^{(\mathrm{a})s}(t,t_0,\bv)
+
\frac{\di\ln\sqrt{P^{(\mathrm{ad})}_\delta(t,k)}}{\di t}\bigg|_{t=t_0}\gamma_\bk^{(\mathrm{b})s}(t,t_0,\bv).
\eeq
As Ref.~\cite{Amin:2025dtd} discussed, these solutions correspond to pure bulk perturbations to the density and velocity at the initial time $t=t_0$. Moreover, these perturbations affect the distribution of every particle $s$ equally.

\paragraph{Sourced solution:}
Finally, the inhomogeneous solution to equation~\eqref{eq:gss_evo} is
\beq\label{eq:gss_iso}
&g^{ss'}_{\bk\bk'}(t,\bv,\bv')
\\&=
4\pi G\bar\rho\,P^{(\mathrm{iso})}_0
\int_{t_0}^t \di t' \left[
\gamma^{(\mathrm{b})s}_{\bk}(t,t',\bv)\gamma^{(\mathrm{c})s'}_{\bk'}(t,t',\bv')
+\gamma^{(\mathrm{c})s}_{\bk}(t,t',\bv,)\gamma^{(\mathrm{b})s'}_{\bk'}(t,t',\bv')
\right]\ddelta(\bk+\bk'),
\eeq
where we define
\beq
P^{(\mathrm{iso})}_0 = \frac{\overline{m^2}}{\overline{m}\,\bar\rho}.
\eeq
Here $\bar\rho$ is the total mass density. We will see that $P^{(\mathrm{iso})}_0$ is the power spectrum of the total density contrast in the absence of correlations.
For adiabatic initial conditions, the two-particle correlation function is the sum of equation~\eqref{eq:gss_ad} and equation~\eqref{eq:gss_iso}.

\subsection{Specialization to particle species}

So far we have considered the probability distributions of individual particles $s$, $s'$, and so on. However, in the limit of an infinite volume, it makes little sense to consider the contribution of each individual particle. Therefore, to make the description more concrete, we may consider subsets $S$, $S'$, $...$ of the particles, which we regard to be ``species''. For simplicity, let us also assume that every particle in a species has the same mass and the same velocity distribution, i.e., $m_s=m_S$ and $f^s(\bv)\propto f^S(\bv)$ for all $s\in S$. No generality is lost since the number of different species can be arbitrarily large. We will nevertheless assume that the number of particles of each species is large, which is always appropriate in the large-volume limit. We define $\mathfrak{f}_S=\bar\rho_S/\bar\rho$ to be the mass fraction in species $S$, where $\bar\rho_S$ is the mass density in species $S$. Note that in these terms
\beq
P^{(\mathrm{iso})}_0
= \frac{1}{\bar\rho}\sum_S m_S \mathfrak{f}_S
= \sum_S \frac{\mathfrak{f}_S^2}{\bar n_S},
\eeq
where $\bar n_S$ is the number density of particles of species $S$.

For convenience, let us normalize our ``species'' distribution function so that it integrates to the total spatial volume, corresponding to $\int\di\bv f^S(\bv)=1$. This requires\footnote{Recall that each particle in $S$ has the same distribution function, so the sum over particles in equations \eqref{eq:fSdef} and~\eqref{eq:gSSdef} is equivalent to multiplication by the number of particles belonging to the species. This factor comes entirely from the normalization convention and should not be interpreted as addition of probabilities.}
\beq\label{eq:fSdef}
f^S(\bv)=\bar n_S^{-1}\sum_{s\in S}f^s(\bv).
\eeq
Note that $f^S$ has dimensions of $(\text{velocity})^{-3}$.
Now for species $S$ and $S'$, the 2-particle distribution function is
\beq\label{eq:gSSdef}
g_{\bk\bk'}^{SS'}(t,\bv,\bv') = \bar n_S^{-1}\bar n_{S'}^{-1} \sum_{s\in S}\sum_{s'\in S'} g_{\bk\bk'}^{ss'}(t,\bv,\bv'),
\eeq
which has dimensions of $(\text{length})^6(\text{velocity})^{-6}$.
By appropriately summing over equations \eqref{eq:gss_ad} and~\eqref{eq:gss_iso}, we obtain
\beq\label{eq:gSS}
&g^{SS'}_{\bk\bk'}(t,\bv,\bv')
\\&=
P^{(\mathrm{ad})}_\delta(t_0,k)\gamma_\bk^{(\mathrm{ad})S}(t,t_0,\bv)\gamma_{\bk'}^{(\mathrm{ad})S'}(t,t_0,\bv')\ddelta(\bk+\bk')
\\&\hphantom{=}
+
4\pi G\bar\rho\,P^{(\mathrm{iso})}_0\int_{t_0}^t \di t' \left[
\gamma^{(\mathrm{b})S}_{\bk}(\bv,t,t')\gamma^{(\mathrm{c})S'}_{\bk'}(t,t',\bv')
+\gamma^{(\mathrm{c})S}_{\bk}(t,t',\bv)\gamma^{(\mathrm{b})S'}_{\bk'}(t,t',\bv')
\right]\ddelta(\bk+\bk')
\eeq
for adiabatic initial conditions, where we define
\beq\label{eq:gammaS}
&\gamma_\bk^{(\mathrm{a})S}(\bv)=f^S(\bv),
\qquad
\gamma_\bk^{(\mathrm{b})S}(\bv)=-\frac{i\bk}{k^2}\cdot\nabla_{\bv} f^S(\bv),
\qquad
\gamma_\bk^{(\mathrm{c})S}(\bv)=\frac{m_S}{\bar\rho\,P^{(\mathrm{iso})}_0} f^S(\bv),
\\
&\gamma^{(i)S}_{\bk}(t,t_0,\bv)
=e^{-i\bk\cdot\bv (t-t_0)}\gamma^{(i)S}_{\bk}(\bv)
+4\pi G\bar\rho\int_{t_0}^t dt'e^{-i\bk\cdot\bv (t-t')} \gamma_\bk^{(\mathrm{b})S}(\bv)\sum_{S'}\int \di\bv'\mathfrak{f}_{S'}\gamma^{(i)S'}_{\bk}(t',t_0,\bv'),
\\
&\gamma_\bk^{(\mathrm{ad})S}(t,t_0,\bv)
=
\gamma_\bk^{(\mathrm{a})S}(t,t_0,\bv)
+
\frac{\di\ln\sqrt{P^{(\mathrm{ad})}_\delta(t,k)}}{\di t}\bigg|_{t=t_0}\gamma_\bk^{(\mathrm{b})S}(t,t_0,\bv).
\eeq

\subsection{The matter power spectrum}

We next note how the matter power spectrum is set by the two-particle correlation function $g^{SS'}$. The mass density of species $S$ is
\beq
\rho_S(t,\bk) = \sum_{s\in S} m_s\, e^{-i \bk\cdot\bx_s}
\eeq
in Fourier space. Now for two species $S$ and $S'$, we define the cross power spectrum $P^{SS'}_\delta(t,k)$ (for $k>0$) by $\bar\rho^2 P_\delta^{SS'}(t,k)\ddelta(\bk+\bk')=\langle \rho_S(t,\bk)\rho_{S'}(t,\bk')\rangle$,
implying that
\beq
\bar\rho^2 P_\delta^{SS'}(t,k)\ddelta(\bk+\bk')
&=
\delta_{SS'}\sum_{s\in S} m_s^2 \langle e^{-i (\bk+\bk')\cdot\bx_s}\rangle
+
\sum_{s\in S}\sum_{s'\in S',\, s'\neq s} m_s m_{s'} \langle e^{-i \bk\cdot\bx_s}e^{-i \bk'\cdot\bx_{s'}}\rangle,
\eeq
where $\delta_{SS'}$ is the Kronecker delta.\footnote{By this definition, $P_\delta^{SS'}$ is the cross power spectrum of $\mathfrak{f}_S\delta_S$ and $\mathfrak{f}_{S'}\delta_{S'}$, where $\delta_S$ is the density contrast of species $S$ and $\delta_{S'}$ is that of $S'$. The cross power spectrum of $\delta_S$ and $\delta_{S'}$ would be $\mathfrak{f}_S^{-1}\mathfrak{f}_{S'}^{-1}P_\delta^{SS'}$.}
But the ensemble averages evaluate to
\beq
\langle e^{-i (\bk+\bk')\cdot\bx_s}\rangle
=
\int\di\bx\di\bv e^{-i (\bk+\bk')\cdot\bx}f^s(\bv)
=
V^{-1}\ddelta(\bk+\bk'),
\eeq
where $V$ is the (arbitrarily large) spatial volume under consideration, and
\beq
\langle e^{-i \bk\cdot\bx_s}e^{-i \bk'\cdot\bx_{s'}}\rangle
&=
\int\di\bx\di\bx'\di\bv\di\bv' e^{-i \bk\cdot\bx}e^{-i \bk'\cdot\bx'} g^{ss'}(t,\bx,\bx',\bv,\bv')
=
\int\di\bv\di\bv' g_{\bk\bk'}^{ss'}(t,\bv,\bv').
\eeq
Consequently, the species cross power spectrum is given by
\beq
P^{SS'}_\delta(t,k)\ddelta(\bk+\bk')
&=
\frac{\frak{f}_S^2}{\bar{n}_S} \ddelta(\bk+\bk')\delta_{SS'}
+
\frak{f}_S \frak{f}_{S'}\int\di\bv\di\bv' g_{\bk\bk'}^{SS'}(t,\bv,\bv').
\eeq

\paragraph{The species power spectrum:}
For adiabatic initial conditions, equation~\eqref{eq:gSS} implies
\beq\label{eq:PSS}
P_\delta^{SS'}(t,k)
&=
\frak{f}_S \frak{f}_{S'}P^{(\mathrm{ad})}_\delta(t_0,k)T_k^{(\mathrm{ad})S}(t,t_0)T_{k}^{(\mathrm{ad})S'}(t,t_0)
+
\frac{\frak{f}_S^2}{\bar{n}_S}\delta_{SS'}
\\
&\hphantom{=}+
4\pi G\bar\rho\,\frak{f}_S \frak{f}_{S'}P^{(\mathrm{iso})}_0 \int_{t_0}^t \di t' \left[
T^{(\mathrm{b})S}_{k}(t,t')T^{(\mathrm{c})S'}_{k}(t,t')
+T^{(\mathrm{c})S}_{k}(t,t')T^{(\mathrm{b})S'}_{k}(t,t')
\right],
\eeq
where we define
\beq
T_k^{(i)S}(t,t_0)
&=
\int\di\bv \gamma_\bk^{(i)S}(t,t_0,\bv).
\eeq
Note that $T^{(\mathrm{a})S}$ and $T^{(\mathrm{c})S}$ are dimensionless, while $T^{(\mathrm{b})S}$ has dimensions of $\text{time}$.
From equation~\eqref{eq:gammaS}, the $T$ satisfy
\beq
T^{(i)S}_{k}(t,t_0)
=
T^{\mathrm{fs}(i)S}_{k}(t,t_0)
&+4\pi G\bar\rho\int_{t_0}^t dt'T_k^{\mathrm{fs}(\mathrm{b})S}(t,t')\sum_{S'}\mathfrak{f}_{S'}T^{(i)S'}_{k}(t',t_0),
\eeq
where we now define
\beq
T^{\mathrm{fs}(i)S}_{k}(t,t_0)
&=
\int\di\bv e^{-i\bk\cdot\bv (t-t_0)}\gamma^{(i)S}_{\bk}(\bv).
\eeq
Note that 
\beq\label{eq:TS_ad}
T_k^{(\mathrm{ad})S}(t,t_0)
&=
T_k^{(\mathrm{a})S}(t,t_0)
+
\frac{\di\ln\sqrt{P^{(\mathrm{ad})}_\delta(t,k)}}{\di t}\bigg|_{t=t_0}T_k^{(\mathrm{b})S}(t,t_0).
\eeq

\paragraph{The total power spectrum:}
The total matter power spectrum is the sum of cross spectra,
\beq
P_\delta(t,k) = \sum_S \sum_{S'}P_\delta^{SS'}(t,k).
\eeq
We can simplify the expression for the total power spectrum by defining new ``total'' transfer functions
\beq
T^{(i)}_k(t,t_0)=\sum_S \mathfrak{f}_S T^{(i)S}_{k}(t,t_0)
\eeq
(which have the same dimensions as the species versions). Then the matter power spectrum is
\beq\label{eq:P_total}
P_\delta(t,k)
&=
P^{(\mathrm{ad})}_\delta(t_0,k)\left[T_k^{(\mathrm{ad})}(t,t_0)\right]^2
+
P^{(\mathrm{iso})}_0
+
8\pi G\bar\rho\,P^{(\mathrm{iso})}_0 \int_{t_0}^t \di t' 
T^{(\mathrm{b})}_k(t,t')T^{(\mathrm{c})}_{k}(t,t'),
\eeq
and these transfer functions satisfy
\beq
T^{(i)}_{k}(t,t_0)
=
T^{\mathrm{fs}(i)}_{k}(t,t_0)
&+4\pi G\bar\rho\int_{t_0}^t dt'T_k^{\mathrm{fs}(\mathrm{b})}(t,t')T^{(i)}_{k}(t',t_0),
\eeq
with
\beq
T^{\mathrm{fs}(i)}_{k}(t,t_0)
&=
\int\di\bv e^{-i\bk\cdot\bv (t-t_0)}\sum_S\mathfrak{f}_S\gamma^{(i)S}_{\bk}(\bv).
\eeq
Note that equation~\eqref{eq:P_total} validates our original definition of $P^{(\mathrm{iso})}_0$. Also
\beq\label{eq:T_ad}
T_k^{(\mathrm{ad})}(t,t_0)
&=
T_k^{(\mathrm{a})}(t,t_0)
+
\frac{\di\ln\sqrt{P^{(\mathrm{ad})}_\delta(t,k)}}{\di t}\bigg|_{t=t_0}T_k^{(\mathrm{b})}(t,t_0).
\eeq

\subsection{Translation to an expanding universe}\label{App:Expanding}

Let $t\rightarrow \eta$ and $G\bar{\rho}\dl t\rightarrow G\bar{\rho} a(\eta)\dl\eta$, where $\dl \eta=\dl t/a^2(t)$ and after the translation, $\bar{\rho}$ and $\bar{n}_S$ are time-independent co-moving densities. Then the total matter power spectrum becomes
\beq
    \label{eq:final}
    P_{\delta}(\eta, k)
    = P^{(\rm ad)}_{\delta}(\eta_0,k) \ml[T^{(\mathrm{ad})}_k(\eta,\eta_0)\mr]^2 + P_{\delta}^{(\mathrm{iso})}(\eta_0, k) \left[1 \!+\! 3\bar{H}_0^2\!\!\int_{\eta_0}^\eta \!\!\dl\eta' a(\eta') T_k^{(\mathrm{b})}(\eta,\eta') {T}_k^{(\mathrm{c})}(\eta,\eta')\right]\!,
\eeq
where $\bar{H}_0^2=(8\pi G/3) \bar{\rho}$, $P^{(\rm iso)}_\delta(\eta_0,k)=\sum_S \mathfrak{f}_S^2/\bar{n}_S$. 
The species power spectra have analogous expressions.
The relevant transfer functions $T^{(\mathrm{ad},\mathrm{b},\mathrm{c})}_k$ can be obtained by solving the following Volterra equation (with $i=\mathrm{ad,b,c})$:
\beq
    \label{eq:SumTab}
    T_k^{(i)}(\eta,\eta_0) &= {T}^{\mathrm{fs}\,(i)}_k(\eta,\eta_0)
    + \frac{3\bar{H}_0^2}{2}\int_{\eta_0}^\eta \dl\eta' a(\eta'){T}^{\mathrm{fs}\,(\mathrm{b})}_k(\eta,\eta') T_k^{(i)}(\eta',\eta_0)\,,
\eeq
where
\beq
    &{T}^{\mathrm{fs}\,(i)}_k(\eta,\eta')=\int_\bv \sum_S\mathfrak{f}_S \gamma_\bk^{S\,(i)}(\bv).\\
\eeq
For the purpose of comparison with simulations and eventually observational data, it is convenient to shift to $y=a/a_{\mathrm{eq}}$ as the time variable. In a universe with matter and radiation, $\eta$, $y$, and the cosmic time $t$ are connected via
\beq
    \label{eq:etaDef}
    \frac{\dl t}{a^2(t)} = \dl\eta =\frac{\sqrt{2}}{a_{\mathrm{eq}}k_{\mathrm{eq}}}\frac{\dl y}{y\sqrt{1+y}},
    \quad \text{with} \quad
    y=\frac{a}{a_{\mathrm{eq}}}.
\eeq
In the main body of the paper, we express our results using $y$ as the independent variable, taking $T_k^{(i)}(\eta,\eta_0)\rightarrow \mathcal{T}_k^{(i)}(y,y_0)$. The results are summarized in section~\ref{sec:Main}.

\section{Perturbative Solutions}
\label{App:Perturbative}
Consider the Volterra equation 
\beq
\mTi(y,y')=\mTfi(y,y')+\frac{3}{2}\int_{y'}^y \frac{\dl y''}{\sqrt{1+y''}} \mTf{b}{k}(y,y'')\mTi(y'',y').
\eeq
Let $\mTi={}^0\mTi+\delta\mTi$ where ${}^0\mTi$ on the right-hand side is the solution  which is zeroth order in $\mathfrak{f}_2$. Similarly $\mTfi={}^0\mTfi+\delta\mTfi$. Then,
\beq
{}^0\mTi(y,y')&={}^0\mTfi(y,y')+\frac{3}{2}\int_{y'}^y \frac{\dl y''}{\sqrt{1+y''}} {}^0\mTf{b}{k}(y,y''){}^0\mTi(y'',y')\,,\\
\delta \mTi(y,y')&=\frac{3}{2}\int_{y'}^y \frac{\dl y''}{\sqrt{1+y''}} {}^0\mTf{b}{k}(y,y'')\delta\mTi(y'',y')\\
\quad &+\underbrace{\delta \mTfi(y,y')+\frac{3}{2}\int_{y'}^y \frac{\dl y''}{\sqrt{1+y''}} \delta\mTf{b}{k}(y,y''){}^0\mTi(y'',y')}_{{\rm source}\,=\,\mathcal{S}^{(i)}_k(y,y')}.
\eeq
Assuming that the zeroth order solution ${}^0\mTi$, as well as $\delta \mTfi$, are known functions, we can solve the ``sourced" equation for $\delta\mTi$. 

\paragraph{Case 1:} Let us first consider the case where the zeroth order solution is that of CDM. Then ${}^0\mTf{b}{k}=\mathcal{F}$, and we get a sourced Mezaros equation:
\beq
(\delta \mTi-\mathcal{S}^{(i)}_k)''+\frac{(2+3y)}{2y(1+y)}(\delta \mTi-\mathcal{S}^{(i)}_k)'-\frac{3}{2y(1+y)}(\delta \mTi-\mathcal{S}^{(i)}_k)=\frac{3}{2y(1+y)}{\mathcal{S}}^{(i)}_k.
\eeq
With  $\delta \mTi-\mathcal{S}^{(i)}_k=0$ initially, the solution is  given by:
\beq
\delta \mTi(y,y')=\mathcal{S}^{(i)}_k(y,y')+\frac{3}{2}\int_{y'}^y \frac{dy''}{\sqrt{1+y''}} {}^{0}\mT{b}{k}(y,y'')\mathcal{S}^{(i)}_k(y'',y').
\eeq
To evaluate the above integral, we need the source function explicitly, which in turn relies of $\delta \mTfi$. For case 1, these are
\beq
\delta \mTf{a}{k}=-\mathfrak{f}_2(1-e^{-\alpha_{k\,2}^2\mathcal{F}^2/2}),\quad \delta \mTf{b}{k}=\mathcal{F}\delta \mTf{a}{k},\quad \delta \mTf{c}{k}= e^{-\alpha_{k\,2}^2\mathcal{F}^2/2}.
\eeq
For the adiabatic part we need 
\beq
\mathcal{S}^{(\mathrm{a})}_k(y,y') &=-\mathfrak{f}_2
\bigg[
(1-e^{-\alpha_{k\,2}^2\mathcal{F}^2(y,y')/2})
\\&\hphantom{=-\mathfrak{f}_2\bigg[}
+\frac{3}{2}\int_{y'}^y \frac{dy''}{\sqrt{1+y''}}(1-e^{-\alpha_{k\,2}^2\mathcal{F}^2(y,y'')/2})\mathcal{F}(y,y''){}^{0}\mT{a}{k}(y'',y')
\bigg]\\
\mathcal{S}^{(\mathrm{b})}_k(y,y') &=-\mathfrak{f}_2
\bigg[
(1-e^{-\alpha_{k\,2}^2\mathcal{F}^2(y,y')/2})\mathcal{F}(y,y')
\\&\hphantom{=-\mathfrak{f}_2\bigg[}
+\frac{3}{2}\int_{y'}^y \frac{dy''}{\sqrt{1+y''}}(1-e^{-\alpha_{k\,2}^2\mathcal{F}^2(y,y'')/2})\mathcal{F}(y,y''){}^{0}\mT{b}{k}(y'',y')
\bigg],
\eeq
which can then be used in the expression for $\delta\mTi(y,y')$. In the limit that $\alpha_{k\,2}\rightarrow 0$, these source functions vanish, and so do the $\delta\mTi(y,y')$. On the other hand, when $\alpha_{k\,2}\rightarrow \infty$, we can evaluate the above integrals analytically. While the detailed expressions are long, in the limit $y\gg1$, $k\gg k_{\rm J}^{\rm eq}$ and $\mathfrak{f}_2\ll 1$, we find
\beq
\mT{ad}{k}(y,y_0)\approx {}^0\mT{ad}{k}\left[1- \frac{1}{5}\mathfrak{f}_2\ml(7.8+3\ln y\mr)\right].
\eeq
The numerical co-efficients are obtained by Taylor expanding the result around $y\rightarrow \infty$.

\paragraph{Case 2:} Now consider the case where the zeroth order solution is that of warm dark matter. While no analytic solution exists for the dominant species, we can still use the same procedure employed above to obtain corrections perturbatively in $\mathfrak{f}_2$. The results are provided in the main body of the paper.

\section{Numerical Algorithm for Power Spectra}
\label{App:Num-PS}
The total power spectrum evolution relies on solving the Volterra equations for $\mathcal{T}_k^{(i)}(y,y_0)$ \eqref{eq:Ty} for $i=\mathrm{a,b,c}$. This is done via an iterative procedure, with the algorithm described in Appendix B of \cite{Amin:2025dtd}. The only difference is that the free streaming kernels we begin with are weighted sums over all the species. 

For the inter- and intra-species power spectra for $N$ species, the $3N$ transfer functions, $\mathcal{T}^{(i)\,S}_k$, satisfy Volterra equation \eqref{eq:Tis}, which is solved using the same iterative procedure mentioned above. Here, the Volterra equation for each $S$ and $i$ is iterated separately. The free-streaming kernels for each species (defined below \eqref{eq:Tis}) are assigned to be the zeroth iteration of the transfer functions. Importantly, the $l$th iteration for the transfer functions is sourced by the weighted sum of the $(l-1)$th iterations of the same type of transfer function for all species appearing in the source term (see \eqref{eq:Tis}).

For $y_0=10^{-3}$ and $y=10^2$, we found that $\sim 15$ iterations are sufficient for convergent results. Discretizing the time ($y$ variable) into $400$ log-spaced intervals (necessary for evaluating intergrals over time), and evaluating the total density spectra at $60$ wavenumbers takes $\sim$ 1 sec in {\it Mathematica 14.0} on a modern laptop. The corresponding algorithm and code can be found at \url{https://github.com/mustafaaamin/warm-and-random.git}.

\end{document}